\documentclass[a4paper,fleqn]{cas-sc}

\usepackage[authoryear]{natbib}
\usepackage{array}
\usepackage{booktabs}
\usepackage{listings}
\usepackage{multirow}
\usepackage{pdflscape}
\usepackage{soul}
\usepackage{svg}
\usepackage{subcaption}
\usepackage{graphicx}
\usepackage{tcolorbox}

\definecolor[named]{BoxColor}{RGB}{226,233,241}
\definecolor[named]{BoxColorTitle}{RGB}{16,78,139}

\newsavebox\CBox
\def\textBF#1{\sbox\CBox{#1}\resizebox{\wd\CBox}{\ht\CBox}{\textbf{#1}}}

\makeatletter
\g@addto@macro{\endtabular}{\gdef\rowfonttype{}}
\makeatother
\newcommand{\rowfonttype}{}
\newcommand{\rowfont}[1]{
   \noalign{\gdef\rowfonttype{#1}}}%
\newcolumntype{R}{>{\rowfonttype\strut}r}

\begin{document}

\title [mode = title]{Enhancing Automatic Keyphrase Labelling with Text-to-Text Transfer Transformer (T5) Architecture: A Framework for Keyphrase Generation and Filtering}

\shorttitle{A Framework for Keyphrase Generation and Filtering}
\shortauthors{Jorge Gabín et~al.}

\author[lkn,irlab]{Jorge Gabín}[auid=000,bioid=1,orcid=0000-0002-5494-0765]
\cormark[1]
\ead{jorge.gabin@udc.es}
\cortext[cor1]{Corresponding author}
\credit{Conzeptualization, Software, Investigation, Writing - Original Draft, Visualization}

\author[lkn]{M. Eduardo Ares}[auid=000,bioid=2,orcid=]
\ead{eduardo@linknovate.com}
\credit{Writing - Review \& Editing}

\author[irlab]{Javier Parapar}[auid=000,bioid=3,orcid=0000-0002-5997-8252]
\ead{javier.parapar@udc.es}
\credit{Conzeptualization, Writing - Review \& Editing, Supervision}

\affiliation[lkn]{organization={Linknovate Science},
            city={A Coruña},
            country={Spain}
}

\affiliation[irlab]{organization={IRLab, CITIC, Computer Science Department, University of A Coruña},
            city={Santiago de Compostela},
            country={Spain}
}

\begin{abstract}
Automatic keyphrase labelling stands for the ability of models to retrieve words or short phrases that adequately describe documents' content. Previous work has put much effort into exploring extractive techniques to address this task; however, these methods cannot produce keyphrases not found in the text. Given this limitation, keyphrase generation approaches have arisen lately. This paper presents a keyphrase generation model based on the Text-to-Text Transfer Transformer (T5) architecture. Having a document's title and abstract as input, we learn a T5 model to generate keyphrases which adequately define its content. We name this model \texttt{docT5keywords}. We not only perform the classic inference approach, where the output sequence is directly selected as the predicted values, but we also report results from a majority voting approach. In this approach, multiple sequences are generated, and the keyphrases are ranked based on their frequency of occurrence across these sequences. Along with this model, we present a novel keyphrase filtering technique based on the T5 architecture. We train a T5 model to learn whether a given keyphrase is relevant to a document. We devise two evaluation methodologies to prove our model's capability to filter inadequate keyphrases. First, we perform a binary evaluation where our model has to predict if a keyphrase is relevant for a given document. Second, we filter the predicted keyphrases by several AKG models and check if the evaluation scores are improved. Experimental results demonstrate that our keyphrase generation model significantly outperforms all the baselines, with gains exceeding 100\% in some cases. The proposed filtering technique also achieves near-perfect accuracy in eliminating false positives across all datasets.
\end{abstract}


\maketitle

\section{Introduction}
\label{sec:intro}

Keyphrase labelling~\cite{xie2023statistical,nasar2019textual}, commonly alluded to as keyphrase extraction or generation, refers to selecting phrases that capture the most important topics in a document. There are two main reasons behind the gain in popularity of keyphrase labelling models in recent years. The first one is the vast amount of documents on the web needing proper descriptors, which dwarfs the human capacity to do manual annotation. News articles, scientific and technical literature, users' topical interests, or social media posts are good examples. Second, the ability of current search engines to exploit that information to produce better rankings, categorise results or even produce summarised representations of the content of the documents. In fact, several information retrieval tasks, such as summarisation~\cite{d2005keyphrase}, recommendation~\cite{ferrara2011keyphrase}, query refinement~\cite{chandrasekar2006system}, exploratory search~\cite{marchionini2006exploratory}, query expansion~\cite{efthimiadis1996query,song2006keyphrase}, document clustering and classification~\cite{steinbach2000comparison,li1998classification} or semantic and faceted search~\cite{guha2003semantic,tunkelang2009faceted}, highly take advantage of these keyphrases (often referred to as keywords). 

Traditionally, researchers have approached the labelling procedure in an extractive way~\cite{papagiannopoulou2020review,hasan2014automatic,vega2019multi,hassan2022lvtia,yan2024utilizing,xhang2023improving,papagiannopoulou2018local}, i.e. the designed models aim to extract words or short phrases from the target text describing the document's content. Nevertheless, keyphrase extraction models have one main disadvantage: they can only label documents with words or phrases that appear in the input text. Thus, keyphrase generation~\cite{ccano2019keyphrase} models emerge as an alternative to the aforementioned models, identifying keyphrases not present in the target documents.

Many information processing tasks depend highly on the existence and the quality of topical annotations of content. For instance, academic search~\cite{10.1145/3038912.3052558}  and recommendation~\cite{10.1145/2505515.2505705} engines, product review exploitation models~\cite{TsurRappopor2009}, or topic detection trend tracking~\cite{panem2014entity} are obvious examples. One indicator of the reliance on keyphrases in academic search and recommendation is that publishers often ask authors to label their publications with keyphrases manually. Those annotated keyphrases are arguably why academic information is the most common domain in keyphrase labelling research.

Large pre-trained models~\cite{han2021pre} have become common and, in many cases, state-of-the-art approaches in many Natural Language Processing (NLP) tasks. These models, along with transfer learning~\cite{pan2009survey}, can adapt the general knowledge learned in the pre-train phase to specific downstream tasks via fine-tuning. One approach gaining massive popularity in the NLP field is T5. Text-to-Text Transfer Transformer (T5)~\cite{raffel2019exploring} is a single framework that, rather than following a multi-step approach, turns all text-based language problems into a text-to-text format.

In this paper, we further develop our previous work on how to apply T5 to the automatic keyphrase generation task~\cite{gabin2022exploring}. We devise a downstream task adapting the keyphrase generation problem to a text-to-text format. We named the resulting model from fine-tuning T5 on this task \texttt{docT5keywords}.

In addition to the keyphrase generation model, and to reduce the number of inadequate keyphrases produced by the proposed generation models, we present a novel keyphrase filtering method that further leverages the T5 architecture. For that, we designed a downstream task to fine-tune T5 to determine whether a keyphrase is relevant to a document. Here we follow the lead of Nogueira et al.~\cite{DBLP:conf/emnlp/NogueiraJPL20} when building their query-document re-ranking model (\texttt{monoT5}). 

We evaluated our proposals by comparing them against a heterogeneous set of state-of-the-art baselines containing supervised and unsupervised Automatic Keyphrase Extraction (AKE) and Automatic Keyphrase Generation (AKG) models. Furthermore, we evaluate our models on many of the most known keyphrase labelling datasets. Details about the baselines and the datasets used are shown in Section~\ref{sec:exp_setup}.

The main contributions of this work are: (1) a thorough analysis of the ability of the T5 architecture to deal with the documents' automatic keyphrase generation task. We compare our model against a wide variety of AKE and AKG models using the most popular datasets for the task. (2) A majority voting-based inference strategy where multiple sequences are generated, and the keyphrases are ranked based on their frequency of occurrence across these sequences. (3) A novel T5-based keyphrase filtering model that may be plugged for keyphrase candidate refinement over the output of any existing model beyond our AKG proposal model.

\section{Research Objectives}
This section outlines the main research objectives of this work.

\begin{enumerate}
    \item \textbf{Study the performance of the T5 generative model}: The first objective is to analyse how the T5 model performs on the keyphrase labelling task, specifically for both exact and partial match evaluations, considering both present and absent keyphrases. This allows us to understand its effectiveness across different types of keyphrases. See Experiments 1 (\S~\ref{sec:e1}) and 2(\S~\ref{sec:e1}), and Appendix~\ref{appendix:partial}. 
    
    \item \textbf{Investigate the impact of incorporating a keyphrase filtering model}: Another important objective is to explore the effect of integrating a filtering model into the pipeline to eliminate irrelevant keyphrases. This study assesses whether including this filtering step improves the quality and relevance of the generated keyphrases. See Experiments 1 (\S~\ref{sec:e1}) and 2(\S~\ref{sec:e1}), and Appendix~\ref{appendix:partial}. 
    
    \item \textbf{Examine the effect of majority voting during the inference phase}: We aim to understand how generating multiple keyphrase sequences and ranking them based on their frequency (majority voting) influences model performance. See Experiments 1 (\S~\ref{sec:e1}) and 2(\S~\ref{sec:e1}), and Appendix~\ref{appendix:partial}. 
    
    \item \textbf{Analyse the effect of sorting keyphrases during fine-tuning}: In this objective, we investigate the effect of ordering keyphrases (present followed by absent) during the fine-tuning phase. The goal is to assess how this strategy impacts the model’s ability to generate both types of keyphrases accurately. See Experiment 3 (\S~\ref{sec:e3}). 
    
    \item \textbf{Evaluate the standalone performance of the keyphrase filtering model}: Finally, we conduct a binary evaluation experiment to measure how well the filtering model can identify relevant keyphrases within a document. This evaluation is key for understanding the model's effectiveness in filtering out irrelevant phrases. See Experiment 4 (\S~\ref{sec:e4}). 
\end{enumerate}

\section{Related work}
\label{sec:related_work}
This section reviews the relevant keyphrase generation and extraction techniques; the T5 architecture, to help understand the fine-tuning we performed in this work; and the previous work on using text-to-text models to document re-ranking.

\subsection{Keyphrase Labelling}
Previous efforts on Automatic Keyphrase Labelling~\cite{xie2023statistical,nasar2019textual} can be divided into the earlier works on AKE approaches and the more recent AKG models. The main difference between AKE and AKG methods is that the first only relies on extracting fragments from the input text to pinpoint the keyphrases. In contrast, the second may produce new keyphrases not seen in the target document. Alternatively, we may classify existing approaches as either supervised or unsupervised models. The most common choice in the existing literature is to follow an unsupervised fashion when working on AKE methods, while most AKG models need supervision in terms of labelled data.

Most AKE algorithms follow a two-step approach:  (1) the algorithm selects a set of words or short phrases as ``candidate keyphrases'' from the input text following different heuristics (2) using either supervised or unsupervised techniques, the method selects the most representative segments for the text topics. Typically, the authors frame this second step as a ranking problem. Thus, most AKE models learn and rank which of the extracted keyphrases are relevant.

Hereafter we describe some unsupervised and supervised AKE techniques we used as baselines in this paper.

Graph-based methods are one of the most common alternatives to approach the AKE task. In particular, \cite{mihalcea2004textrank}, with their model named \texttt{TextRank}, were the first authors to approach this task using graph-based techniques. First, their model builds a graph using text units filtered via part-of-speech tags as vertices, and it decides that there is an edge between two vertices if they co-occur in a text window. Then it applies the PageRank algorithm to select the keyphrases. The evolution of this method is \texttt{SingleRank}~\cite{wan2008single}, which improves the former version using weights on the graph's edges. Litvak and Last~\cite{litvak2008graph} also tested using HITS instead of PageRank for keyword-based summarising with good results. Other graph-based methods, such as spectral clustering~\cite{10.1145/564376.564398}, community detection models~\cite{10.1145/1526709.1526798}, or more recently, Graph Convolutional Networks~\cite{10.1145/3331184.3331219}  have also been tested for the task.  

Another commonly followed approach for this task is to use topic-based methods that identify salient keyphrases. One example of this type of model is \texttt{TopicRank}~\cite{bougouin2013topicrank}. First, this technique pre-processes the text and extracts candidate keyphrases. After that, the model clusters the candidates using Hierarchical Agglomerative Clustering~\cite{day1984efficient}. With the resulting clustering, the algorithm builds a graph of topics with weighted edges based on a measure that considers phrases' offset positions in the text. Finally, the model uses \texttt{TextRank} to rank the topics, and then, it selects one candidate keyphrase from each of the top topics. In this line, Zhang et al. presented HTKG~\cite{10.1145/3477495.3531990}, a method based on neural hierarchical topic models for keyphrase generation. 

Some AKE techniques leverage the words' statistical information to select keyphrases. One example is \texttt{KP-Miner}~\cite{el2009kp}. This model uses a reasonably appropriate filtering method to select the candidate keyphrases and then uses a scoring function similar to \textit{tf·idf} to perform the final sampling. A more recent approach using statistical and contextual information is \texttt{YAKE}~\cite{campos2018text}. \texttt{YAKE} both leverages the term's position and frequency and uses novel statistical metrics to capture context information.

Following recent trends in text representations, \texttt{EmbedRank}~\cite{bennani2018simple} exploits embeddings for representing candidate sentences and documents. Then the model uses cosine similarity between candidate keyphrases and documents' embeddings to rank them and choose the top as the final result.  

One popular supervised AKE model is \texttt{KEA}~\cite{witten1999kea}. This system first performs the candidate identification using several lexical methods, computing a set of features for each. Then it uses a Naïve Bayes classifier to predict whether the candidate keyphrases are adequate for the document. 

Unlike AKE systems, keyphrase generation models do not limit the candidates to the words or phrases in the input text. This is also the case with some extraction-based techniques that expand their models with a keyphrase generation module to obtain absent keyphrases. For example, Zahera et al.~\cite{zahera2022multpax} link the present keyphrases to an existing knowledge graph to enrich the extracted keyphrases set with words or phrases not present in the text. However, the most common approach to AKG is using sequence-to-sequence models without depending on an extraction step. 

Although other alternatives exist, sequence-to-sequence models have been postulated as the best approach to the keyphrase generation task. One of the first and most popular models to interpret the AKG task as a sequence-to-sequence problem was \texttt{catSeq}~\cite{DBLP:conf/acl/YuanWMTBHT20}. This model, based on the \texttt{Seq2Seq} framework~\cite{sutskever2014sequence}, concatenates multiple phrases into a single sequence by using a delimiter (\texttt{<sep>}). Then, during the training phase, this concatenated sequence of words or short phrases is used as the target for a sequence generation task. 

Later, Ye et al.~\cite{DBLP:conf/acl/YeGL0Z20} noticed that the \texttt{catSeq} model was conditioned by the keyphrases order, while in real-world applications, they are rather an unordered set than an ordered sequence. With these assumptions, they propose adapting the \texttt{catSeq} model to work with sets instead of sequences, \texttt{One2Set}. This model is based on \texttt{SetTrans} a Transformer-based sequence-to-sequence model presented along with \texttt{One2Set} that can generate sets instead of sequences.

A recent contribution to the AKE and AKG tasks is the approach proposed by Kulkarni et al.~\cite{kulkarni2022learning}. Their work introduces two models, namely \texttt{KBIR} and \texttt{KeyBART}, designed to address the keyphrase extraction and generation tasks, respectively. These models leverage slight adaptations to the BERT architecture and pre-training process to enhance the quality of keyphrase representations. Both \texttt{KBIR} and \texttt{KeyBART} models have demonstrated impressive performance across various keyphrase-related tasks, in addition to excelling in their primary objectives of AKE and AKG, where they are arguably the state-of-the-art.

Lately, due to the increasing popularity of LLM-based chat-bots, like \texttt{ChatGPT} or \texttt{Bard}, researchers have been putting much effort into testing their capabilities on different tasks. One of the tasks these models have been tested on is AKE/AKG. Published works on this matter~\cite{song2023chatgpt,martinez2023chatgpt} show promising results on these tasks.

\subsection{Text-to-Text Transfer Transformer (T5)}
Current massive language models pursue general-purpose knowledge to be able to ``understand'' the text. Because of that, most of the recent proposals pre-train the understanding ability through a task complementary to the target one (e.g. span corruption, token masking). Transfer learning techniques allow domains, tasks, and distributions to differ among training and testing phases. Thus, the base pre-trained models can be fine-tuned on specific downstream tasks, transferring knowledge obtained in the general pre-training phase to the particular task.

Many artificial intelligence fields, such as computer vision or signal processing, rely on supervised transfer learning. However, given the nature of the tasks and data, modern state-of-the-art NLP models favour unsupervised transfer learning instead.

In this context, T5 is a text-to-text model that aims to transform every NLP problem into a text-to-text task. \cite{raffel2019exploring} present T5 as a slight adaptation of the original Transformer architecture~\cite{DBLP:conf/nips/VaswaniSPUJGKP17}. First, they remove the Layer Form bias by using a different embedding position scheme and placing the layer normalisation outside the residual path. Second, T5 uses relative position embeddings, a novel alternative that replaces fixed position embeddings used in the traditional Transformer architecture. This way, the model learns embeddings according to the offset between the keys and queries that the attention mechanism processes. An attention function can be defined as mapping a query and a set of key-value pairs to an output, where the query, keys, values, and output are all vectors.

In~\cite{raffel2019exploring}, the authors introduce the ``Colossal Clean Crawled Corpus'' (C4), a cleaned dataset of web-extracted text from April 2019 provided by Common Crawl. This corpus is utilised for pre-training T5 models through an unsupervised span corruption task. As noted earlier, T5 aims to convert text-processing challenges into text-to-text tasks, demonstrating its effectiveness across various downstream applications.

Recent research aims to improve T5 by slightly adapting its architecture or changing the pre-training and fine-tuning process. For example, \texttt{LongT5}~\cite{guo2022longt5} explores the effects of scaling both input length and model size at the same time. On the other hand, \texttt{FlanT5}~\cite{chung2022scaling} studies how increasing the model size and using more fine-tuning tasks affect the model's performance. 

\subsection{Text-to-Text Re-ranking Models}
Previous works on ranking and re-ranking documents have proven the effectiveness of text-to-text models and, specifically, T5-based models for approaching this task. In~\cite{pradeep2021expando}, Pradeep et al. present Expando-Mono-Duo, a complete framework to address ranking tasks using a query generation model (\texttt{doc2query-T5})~\cite{nogueira2019doc2query} and two re-ranking models (\texttt{monoT5}~\cite{DBLP:conf/emnlp/NogueiraJPL20} and \texttt{duoT5}). All three models are text-to-text and follow the T5 architecture. The Expando-Mono-Duo T5 multi-stage ranking architecture comprises four steps: (1) use the \texttt{doc2query-T5} model to expand the documents with queries, (2) compute the first ranking using BM25, (3) perform point-wise re-ranking using \texttt{monoT5} and (4) perform pair-wise re-ranking using \texttt{duoT5}.

The \texttt{doc2query-T5} model is trained with pairs of document-query from the MS MARCO~\cite{DBLP:conf/nips/NguyenRSGTMD16} dataset to produce relevant queries for a target document. Then, the authors append these queries to the documents performing document expansion resulting in an improved ranking accuracy for ad-hoc retrieval.

On the other hand, the idea behind the \texttt{monoT5} is to use a text-to-text model to determine how relevant a document is for a query. To achieve this, the authors~\cite{DBLP:conf/emnlp/NogueiraJPL20} first fine-tune the model with the following sequence:
\begin{center}
    Query: $q$ \space\space Document: $d$ \space\space Relevant: $r$
\end{center}
\noindent where $q$ and $d$ represent the query and document texts, respectively; and $r$ (the target) is ``\textit{true}'' or ``\textit{false}'' depending on if the query $q$ is relevant or not to the document $d$.

The authors only use a softmax on the logits of the ``\textit{true}'' and ``\textit{false}'' tokens at inference time to measure likelihoods for each query-document pair.

Alternatively, \texttt{duoT5} is a pair-wise re-ranker that aims to determine if a document is more relevant than another document to a given query. Having this in mind, the authors fine-tune a T5 model with the subsequent sequence:
\begin{center}
    Query: $q$ \space\space Document0: $d_i$ \space\space Document1: $d_j$ \space\space Relevant: $r$
\end{center}
where $q$ is the query text, $d_i$ and $d_j$ are two documents' texts; and $r$ (the target) is ``\textit{true}'' if $d_i$ is more relevant then $d_j$, and ``\textit{false}'' otherwise.

To perform the re-ranking at the inference stage, authors try different aggregation methods with the pair-wise scores $p_{i,j}$. This way, every document has a single score $s_i$.

\section{Proposal}
\label{sec:proposal}
Our proposal further explores the text generation capacities of the T5 text-to-text architecture in the keyphrase generation task. We present two complementary T5-based models: a keyphrase generation model and a keyphrase filtering model. We name these models \texttt{docT5keywords} and \texttt{keyFilT5r}, respectively.

\subsection{Base Models}
In this work, all the models we fine-tune have one of these two starting points, \texttt{t5-base} or \texttt{FlanT5}. Here we briefly describe the pre-training and fine-tuning processes followed by the authors of T5 to build these commonly used baseline models.

As previously commented, \texttt{t5-base} is pre-trained in the C4 corpus. The unsupervised pre-training follows the so-called span corruption task that pursues acquiring general-purpose knowledge to apply it to different downstream tasks. The task trains a model to predict the missing parts of a given text. Span corruption is somehow similar to the task used during BERT's pre-training, but instead of one token per masked word, it uses one for each missing part of the text, which can be only one or several words.

One of the objectives of the T5 architecture is to build a model that can address several NLP tasks. With this in mind,  \texttt{t5-base} was further fine-tuned on several common NLP tasks, such as sentence acceptability judgement, sentiment analysis, and sentence completion.

The other base model used in this work is \texttt{FlanT5-base}. This model is the result of further fine-tuning the \texttt{t5-base} model in 1,836 tasks by combining four mixtures from prior work: Muffin, T0-SF, NIV2, and CoT. This fine-tuning approach has shown significant improvements over the base model on several evaluation benchmarks.

Note that most T5-based models have a 512 sequence length limitation in practice due to the computational cost. Therefore, after building the input sequences, in any of our proposals, we keep the first 512 tokens if the limit is exceeded.

\subsection{Tuning T5 for Keyphrase Generation} \label{subsec:akg_train}
\begin{figure}[t]
  \centering
  \includegraphics[width=\columnwidth]{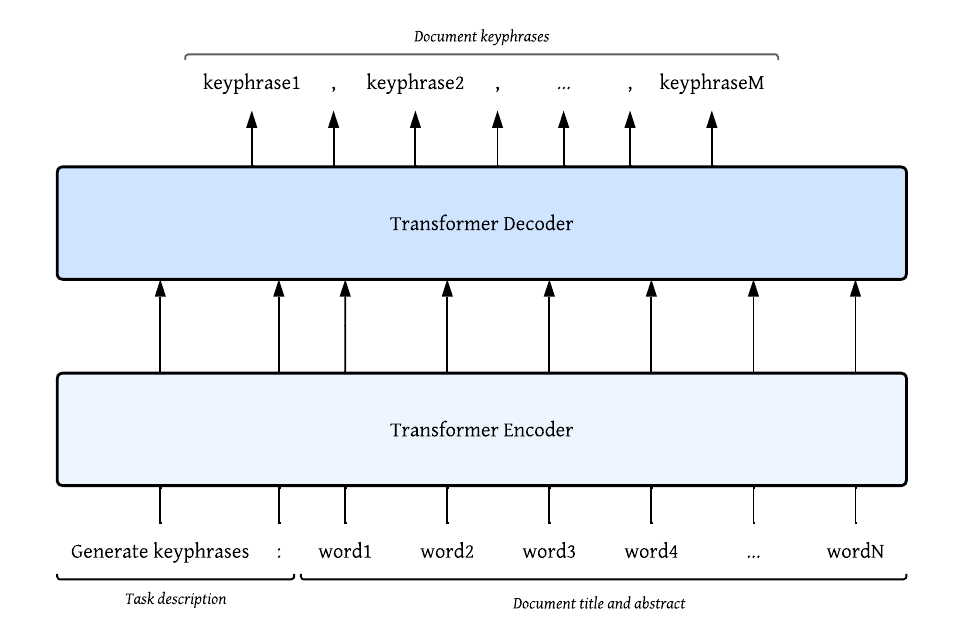}
  \caption{Fine-tuning T5 for keyphrase generation. To anticipate the keyphrases, both the task description and the document tokens are fed into the Transformer encoder.}
  \label{fig:docT5keywords}
\end{figure}

The task we are addressing here aims to deliver a set of keyphrases that adequately represent a document (e.g. scientific publication or news article) based on its textual content. Therefore, to fine-tune a T5 model for keyphrase generation, we must first translate this task into a text-to-text task. Figure~\ref{fig:docT5keywords} shows the fine-tuning setup to train the \texttt{docT5keywords} model. We use a task descriptor and concatenate the document's title and abstract as the input and the document's annotated keyphrases as labels. During the inference phase, we feed the fine-tuned model with documents' titles and abstracts which will output a set of keyphrases for each.

In our proposal, we tested three fine-tuning approaches. The first method (\texttt{vanilla}) fine-tunes the \texttt{t5-base} model for keyphrase generation in the training split of the evaluation dataset. The second one (\texttt{FlanT5}) uses \texttt{flan-t5-base} as the base model instead of \texttt{t5-base}. The third proposal (\texttt{MAG ft}) takes document and keyphrase pairs for fine-tuning from the  Microsoft Academic Graph (\S~\ref{subsec:mag}), an external dataset of annotated scientific publications. Then, we further fine-tune the model in the training split of the evaluation dataset. Section~\ref{sec:exp_setup} provides detailed information about the fine-tuning approaches and the used datasets.

In addition to the three fine-tuning strategies, we also evaluate our models using two different inference approaches. The first approach employs greedy decoding, where the model selects the highest probability output at each step of sequence generation, based on the conditional probability given the preceding tokens. The second approach employs beam-search multinomial sampling to perform majority voting across the generated keyphrases. Instead of producing a single output, we generate one output per beam, count the frequency of each keyphrase across all beams, and then rank them based on their frequency.

Using a generative text-to-text model like \texttt{docT5keywords} for labelling documents makes it possible to identify keyphrases not present in the document's content. Those keyphrases are commonly referred to as ``absent keyphrases'' or, in the context of generative models, ``creative keyphrases''. Producing creative descriptors is a remarkable advantage against most previous models, primarily based on extraction and, therefore, limited to the words or phrases in the input text.

\subsection{Tuning T5 for Keyphrase Filtering}
Creative models have, however, one main disadvantage. It is not uncommon that they hallucinate when generating content. ``Hallucination'' in the Natural Language Generation (NLG) field stands for the problem that models produce text that is entirely not in line with the input data. This issue is well-known in several NLG areas, such as image captioning~\cite{rohrbach-etal-2018-object}. Recent NLP and NLG research indicates that hallucination is a relevant issue in many of these models' applications. Ji et al.~\cite{ji2022survey} review current strategies for mitigating hallucination in diverse domains of NLG.

\begin{figure}[t]
  \centering
  \includegraphics[width=\columnwidth]{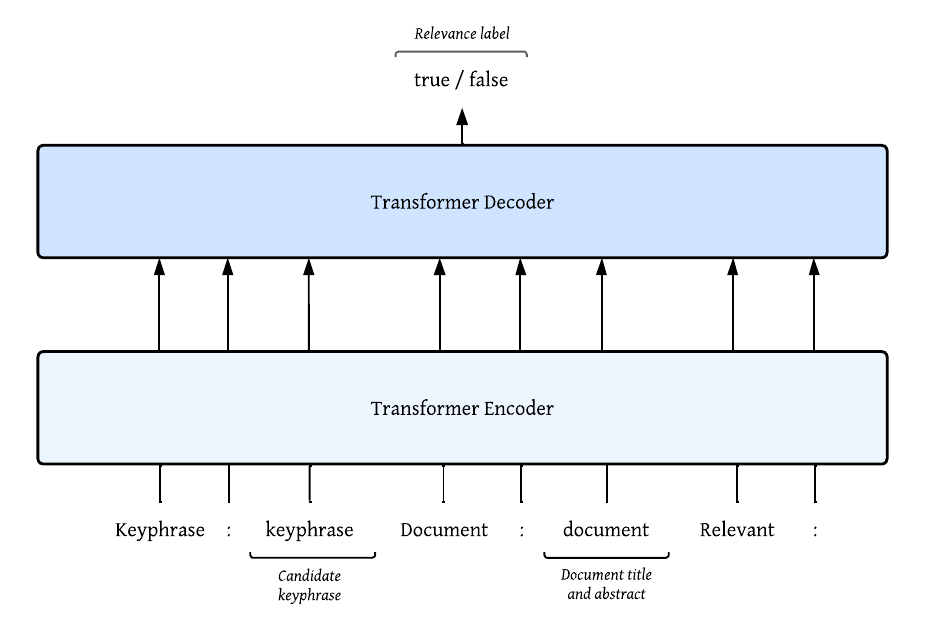}
  \caption{Fine-tuning T5 for keyphrase filtering. Both the keyphrase and the document it refers to are fed into the Transformer encoder along with the task descriptors to classify it as relevant or irrelevant.}
  \label{fig:keyFilT5r}
\end{figure}

Our model may hallucinate by generating keyphrases unrelated to the document's content. For instance, it may generate a list of keyphrases coherent among them, but some of the keyphrases are unrelated to the target input's specifics. Labelling documents with words or phrases that are inadequate for a document may not significantly affect the metrics we use to evaluate the models; nevertheless, it worsens the user experience in a real-world system. Considering this, we develop a keyphrase filtering model based on the \texttt{monoT5} fine-tuning approach. The idea is to filter out unrelated keyphrases from the output of the keyphrase generation model. As shown in Figure~\ref{fig:keyFilT5r}, we devise a fine-tuning task where we input a keyphrase and a document's title and abstract. Then, the model has to predict if the keyphrase is either relevant (``\textit{true}'') or not (``\textit{false}'') to the input document. 

The fine-tuning approach we follow here is equivalent to the one used by Nogueira et al.~\cite{DBLP:conf/emnlp/NogueiraJPL20} to train the \texttt{monoT5} model. In our case, the keyphrases play the role that queries have in \texttt{monoT5}. They define a task to identify if a document is relevant or not to a given query following the same structure that we use. In their case, the model is later used in a document re-ranking task by using the ``\textit{true}'' and ``\textit{false}'' logits to sort the documents retrieved by a query. In our case, we are not interested in re-ranking keyphrases but in eliminating hallucinations for the generated set of keyphrases. Therefore, we take out of the results those generated keyphrases that the filtering model categorises as \textit{false}. 

We design two different training examples generation strategies which we detail in the subsequent sections.

\subsubsection{Soft Negative Sampling}
The first strategy follows the lead of Nogueira et al.~\cite{DBLP:conf/emnlp/NogueiraJPL20}, establishing a 1-to-1 ratio between positive and negative examples. That is, if a training document has 5 annotated keyphrases, we also select 5 negative keyphrases, and thus, we feed the model with 10 different examples. We use keyphrases belonging to the input document as positives. On the other hand, we employ the keyphrases co-occurrence graph's connected components to extract the negative examples. This procedure is similar to the one used in~\cite{gabin2023keyword}. Here, instead of selecting negatives outside the keyphrase's connected component, we use other keyphrases in the same connected component but not from the document's keyphrases set as negative examples. One edge case we must address here is when only the keyphrases belonging to a document compose the connected component; when this happens, we select any keyphrase outside the document's keyphrases set.

We must generate both positive and negative examples to train the filtering model. Following the strategy used in \cite{DBLP:conf/emnlp/NogueiraJPL20}, we establish a 1-to-1 ratio between positive and negative examples. That is, if a training document has 5 annotated keyphrases, we also select 5 negative keyphrases, and thus, we feed the model with 10 different examples. We use keyphrases belonging to the input document as positives. On the other hand, we employ the keyphrases co-occurrence graph's connected components to extract the negative examples. This procedure is similar to the one used in~\cite{gabin2023keyword}. Here, instead of selecting negatives outside the keyphrase's connected component, we use other keyphrases in the same connected component but not from the document's keyphrases set as negative examples. One edge case we must address here is when only the keyphrases belonging to a document compose the connected component; when this happens, we select any keyphrase outside the document's keyphrases set.

\begin{figure}
    \centering
    \includegraphics[width=\columnwidth]{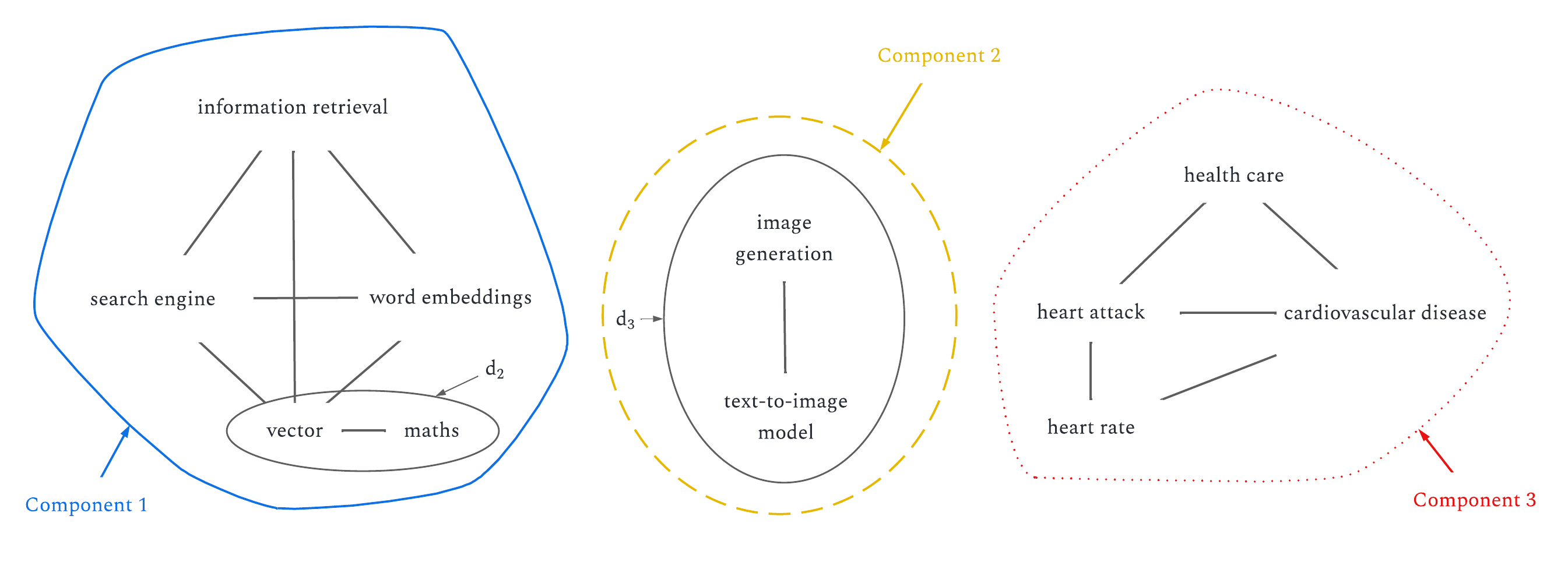}
    \caption{Example of a keyphrase co-occurrence graph with its connected components.}
    \label{fig:con_comps}
\end{figure}

Figure~\ref{fig:con_comps} shows an example of a keyphrase co-occurrence graph and its connected components. In that example, the negative examples of the document $d_2$ will be extracted from the set of keyphrases that are in the same connected component its keyphrases are part of (\textit{Component 1}), but do not belong to $d_2$. Thus, $d_2$'s negative examples would be randomly sampled from the following set of keyphrases: \{\textit{information retrieval, search engine, word embeddings}\}. As previously mentioned, if all the keyphrases of a connected component belong to the same document, we extract the negative examples from a different connected component. For example, $d_3$'s negative examples are extracted from any other document.

Formally, the \texttt{keyFilT5r} training examples for a document $d$ ($TS_d$) are built as follows:

\begin{equation*}
    TS_d = \Bigg\{ (d, k_p, k_{sn}) \;\vert\; k_p \in K_d,\; k_{sn} \in C(k_p) \setminus K_d \Bigg\}
\end{equation*}

\noindent where $k_p$ is a positive keyphrase, $k_{sn}$ the selected negative keyphrase, $K_d$ the set of keyphrases of the document, and $C(k_p)$ the set of keyphrases in the same connected component as $k_p$. In the case where $C(k_p) \setminus K_d$ is $\emptyset$, we select $k_{sn}$ from $K \setminus K_d$, this is all the keyphrases ($K$) except the ones that belong to the document ($K_d$).

\subsubsection{Mixed Negative Sampling}
In addition to the previously explained training examples generation strategy, we propose an alternative method that employs the keyphrase generation model to compute the hard negative examples. This way, we mix both soft and hard negative keyphrases to enhance the filtering model training data. First, we must define train, validation, and test splits for the datasets. We use the training set to learn the keyphrase generation model. Then, we use the validation set to train the \texttt{keyFilT5r} model, leveraging the keyphrases generated by the model for these documents. Note that in the case where the used dataset does not have an explicit validation split, we extract a small portion from the training set. Finally, the test set is the same set we used in the keyphrase generation task evaluation. 

Regarding the keyphrase generation model training, we follow the strategy explained in Section~\ref{subsec:akg_train}. Then, as we advanced earlier, we used this model to generate keyphrases for every document in the validation set. The generated keyphrases not present in the ground truth constitute hard negatives. Concerning the soft negative and the positive examples, we employ the same strategy we used in the previously explained approach.

Formally, these training examples for a document $d$ ($TS_d$) are built as follows:

\begin{equation*}
    TS_d = \Bigg\{ \left(d, k_p, \left\{k_{sn}^1\:...\:k_{sn}^i\right\}, \left\{k_{hn}^1\:...\:k_{hn}^j\right\}\right) \;\Big\vert\; k_p \in K_d,\; k_{sn} \in C(k_p) \setminus K_d,\; k_{hn} \in GK_d \setminus K_d \Bigg\}
\end{equation*}

\noindent where each $k_{hn}$ is a hard negative keyphrase selected from the generated keyphrases set for a specific document ($GK_d$) that do not appear in the document's annotated set of keyphrases, and $i$ and $j$ are the number of soft and hard negative examples per positive one respectively.

\section{Experimental setup}
\label{sec:exp_setup}
This section depicts the datasets used to train and test the models, the baselines used to assess our models' performance, the evaluation methodologies, and the models' parameters selection.

\subsection{Datasets}
Tables~\ref{table:dataset_main} and \ref{table:dataset_splits} show some statistics about the datasets used in this work. Hereafter, we will describe each of them to understand better the training and testing data used in this work.

\begin{table}[tb]
\centering
\caption{Main statistics of the datasets grouped by type.}
\label{table:dataset_main}
\footnotesize
\renewcommand{\arraystretch}{1.5}
\setlength{\tabcolsep}{8pt}
\begin{tabular}
{llrclc}
\toprule
Type & Name & Docs. & $\overline{|K_d|}$ & Annotation type & Absent  ratio \\
\midrule
\multirow{4}*{Abstracts} & Inspec (\cite{hulth2003improved}) & 2,000 & 10 & Indexers & 0.24 \\
 & KP20k (\cite{meng2017deep}) & 553,387 & 5 & Authors & 0.45 \\ 
 & KP-BioMed (\cite{houbre2022large}) & 5,625,688 & 5 & Authors & 0.34 \\
 & MAG (\cite{wang2020microsoft}) & 34,512,382 & 13 & Authors & 0.75 \\ \hline
\multirow{1}*{News} & KPTimes (\cite{DBLP:conf/inlg/GallinaBD19}) & 289,923 & 5 & Semi-automatic & 0.54 \\ 
\bottomrule
\end{tabular}
\end{table}

\begin{table}[tb]
\centering
\caption{Number of documents per split for each evaluation dataset.}
\label{table:dataset_splits}
\footnotesize
\renewcommand{\arraystretch}{1.5}
\setlength{\tabcolsep}{8pt}
\begin{tabular}
{llrrr}
\toprule
Type & Name & Train split & Valid. split & Test split \\
\midrule
\multirow{3}*{Abstracts} & Inspec (\cite{hulth2003improved}) & 1,000 & 500 & 500 \\
 & KP20k (\cite{meng2017deep}) & 514,154 & 19,616 & 19,617 \\ 
 & KP-BioMed (\cite{houbre2022large}) & 5,585,688 & 20,000 & 20,000 \\ \hline
\multirow{1}*{News} & KPTimes (\cite{DBLP:conf/inlg/GallinaBD19}) & 259,923 & 10,000 & 20,000 \\ 
\bottomrule
\end{tabular}
\end{table}

\subsubsection{Inspec}
This collection~\cite{hulth2003improved} consists of 2,000 titles and abstracts from scientific journal articles. Each document has been annotated by professional labellers with two sets of keyphrases: controlled and uncontrolled. The controlled set includes keyphrases found in the Inspec thesaurus, while the uncontrolled set contains any relevant terms based on the document's content. Consistent with the standard practice in AKE and AKG research, we utilise only the uncontrolled keyphrase sets.

\subsubsection{KP20k}
This dataset~\cite{meng2017deep} is a keyphrase labelling collection of 567,830 computer science articles obtained from various online digital libraries such as ACM Digital Library, ScienceDirect, Wiley, and Web of Science. Dataset authors employ papers' titles and abstracts as the source text and only use author-annotated keyphrases.

\subsubsection{KP-BioMed}
This dataset~\cite{houbre2022large} includes approximately 5.6 million titles, abstracts, and author keyphrases extracted from the December 2021 baseline set of MEDLINE/PubMed citation records\footnote{https://ftp.ncbi.nlm.nih.gov/pubmed/baseline}. The collection was randomly and uniformly partitioned by publishing year into training, validation, and test sets. To examine how varying amounts of training data affect keyphrase generation quality, the training set was further split into subsets of different sizes: small (500k), medium (2M), and large (5.6M), with each subset also evenly divided by publishing year.

\subsubsection{MAG}
\label{subsec:mag}
The Microsoft Academic Graph (MAG)~\cite{wang2020microsoft} is a heterogeneous graph encompassing records of scientific publications, citation links, authors, institutions, journals, conferences, and research topics.

We utilised MAG papers from the Open Academic Graph v1 (OAG v1) dataset, which was formed by merging the Microsoft Academic Graph and AMiner. The decision to use OAG v1 over OAG v2 was based on the fact that the MAG papers in OAG v2 did not include author keyphrases.

The OAG v1 dataset comprises 166,192,182 MAG documents. We pre-processed these documents by removing extraneous characters and retaining only those with titles, abstracts, and keyphrases. Documents and keyphrases not in English were also filtered out, resulting in a final collection of nearly 35 million documents and their associated keyphrases.

\subsubsection{KPTimes}
This dataset~\cite{DBLP:conf/inlg/GallinaBD19} is a large-scale collection of news articles annotated with editor-curated keyphrases. The documents in this corpus were annotated in a semi-automatic way. First, editors revise a set of keyphrases provided by an algorithm. Then, they provide additional keyphrases that can be used to improve the algorithm.

\subsection{Baselines}
As introduced in Section~\ref{sec:related_work}, we include several baselines based on different AKE/AKG approaches to produce keyphrases given documents' content. Hereafter we describe how we trained the baseline models when needed, and how we performed inference with each of them. Note that we do not make a literature review, instead, we run all the baselines ourselves trying to reproduce the results presented by their authors.

Regarding the following models: \texttt{TextRank}~\cite{mihalcea2004textrank}, \texttt{SingleRank}~\cite{wan2008single}, \texttt{TopicRank}~\cite{bougouin2013topicrank}, \texttt{YAKE}~\cite{campos2018text} and \texttt{KEA}~\cite{witten1999kea}; we make use of \texttt{pke}~\cite{boudin:2016:COLINGDEMO}, an open source python-based keyphrase extraction toolkit. The first five listed models are unsupervised, so no training is needed. On the other hand, \texttt{KEA} is a supervised model, so we train the model on the training split of each evaluation collection to then execute inference on them.

In the case of \texttt{EmbedRank}~\cite{bennani2018simple}, the remaining unsupervised AKE method, we follow the instructions in the GitHub repository\footnote{https://github.com/swisscom/ai-research-keyphrase-extraction} provided by the model's authors to predict keyphrases.  

The \texttt{One2Set}~\cite{DBLP:conf/acl/YeGL0Z20} model, belongs to the supervised AKG models category. As a supervised model, training it before executing inference is necessary. To do so, we use the scripts provided by its authors in their GitHub repository\footnote{https://github.com/jiacheng-ye/kg\_one2set}. Please note that in \cite{DBLP:conf/acl/YeGL0Z20} the authors trained a single model in the training split of the KP20k. We replicated this training configuration in our work.

One of the most recent and successful approaches to the AKG task is \texttt{KeyBART}~\cite{kulkarni2022learning}. This adaptation of the BERT architecture and pre-training is already pre-trained on the task, but the authors proved that further fine-tuning the model in the AKG task produced better results. Following this advice, we fine-tune the model on the KP20k dataset as model authors do in their work.

Finally, given the arousal and wide use of LLM-based chat-bots like \texttt{ChatGPT} or \texttt{Gemini}, we include a baseline representing these models. For the sake of reproducibility, we use \texttt{Vicuna}~\cite{touvron2023llama,zheng2024judging}, an open-source chat-bot trained by fine-tuning \texttt{LLaMA} on user-shared conversations collected from ShareGPT\footnote{https://sharegpt.com}.

\subsection{Evaluation}
This section describes the different evaluation techniques we used to assess our keyphrase generation models and our keyphrase filtering model performance. For the keyphrase identification task, we employed four evaluation strategies based on whether the extracted/generated keyphrases exactly or partially match the present and absent golden keyphrases.

\subsubsection{Keyphrase Exact Match Evaluation}\label{subsec:em}
The exact match evaluation approach determines if the extracted or generated keyphrases exactly match those in the ground truth. However, this method has a notable drawback: it penalises methods even if they identify semantically equivalent keyphrases. Additionally, direct comparison of model outputs to golden labels is problematic since models might generate keyphrases with slight variations from the ground truth, such as differences in number or verb tense. To mitigate this issue, we employed a stemmer to process both model outputs and reference keyphrases, removing characters like dashes or spaces to account for variations in spelling.

Formally, we can define the set of exact matches for a document ($EM_d$) as follows:
\begin{equation*}
    EM_d = \Big\{p_i \in P_d^s \;\vert\; p_i \in K_d^s\Big\}
\end{equation*}
\noindent where $P_d^s$ is the predicted set of stemmed keyphrases of the document $d$ and $K_d^s$, is the document's true set of stemmed keyphrases.

\subsubsection{Keyphrase Partial Match Evaluation}\label{subsec:pm}
The partial match evaluation does not require a complete match between the produced keyphrases and the ones in the golden truth. Instead, it checks if any of the generated keyphrases is part of any keyphrase of the golden set and vice versa. This method still penalises performance even if keyphrases are semantically equivalent, but it is much less strict than the previous one. Again, we use stemming methods to alleviate syntactical differences between the predictions and the golden labels.

Formally, we can define the set of partial matches for a document ($PM_d$) as follows:
\begin{equation*}
    PM_d = \Big\{p_i \in P_d^s \;\vert\; \exists k_i \in K_d^s\:, (p_i \sqsubseteq k_i) \vee (k_i \sqsubseteq p_i)\Big\}
\end{equation*}
\noindent where $P_d^s$ is the predicted set of stemmed keyphrases of the document $d$, $K_d^s$  is the document's true set of stemmed keyphrases and $p \sqsubseteq k$ evaluates if $p$ is a sub-string of $k$.

\subsubsection{Present Keyphrase Evaluation}\label{subsec:pkps}
Present keyphrases refer to the keyphrases that are actually mentioned or present in the input document or text. The evaluation of present keyphrases involves measuring the ability of a keyphrase extraction or generation model to accurately identify and reproduce the keyphrases that are explicitly stated or described in the input text.

\subsubsection{Absent Keyphrase Evaluation}\label{subsec:akps}
In addition to evaluating present keyphrases, assessing the generation of absent keyphrases is crucial for the AKG task. Absent keyphrases, also known as missing or implicit keyphrases, are the relevant keyphrases that are not explicitly mentioned in the input text but are conceptually related and important for capturing the overall content and context. Evaluating absent keyphrases involves measuring the model's ability to generate relevant keyphrases that are not explicitly present in the input text but are contextually appropriate and provide a complete representation of the content.

\subsubsection{Keyphrase Filtering Evaluation}\label{subsec:filtering_eval}
We follow two strategies to assess the performance of our keyphrase filtering model. The first idea is simple: we filter the keyphrases produced by the different models and compare the exact and partial evaluation metrics between the filtered and unfiltered sets of keyphrases. The second method we use to evaluate the model is a binary evaluation approach. This time, the model has to predict if a keyphrase is relevant or not to a specific document based on its content. The metric used in the binary evaluation is accuracy.

\subsubsection{Evaluation Metrics}
We employ the classical ranking metric, $F_1$ score, which computes the harmonic mean of the precision and recall for the keyphrase labelling task. We compute $F_1$ at 5 and $\mathcal{O}$, being $\mathcal{O}$ the number of golden keyphrases for each document. Note that we do not count matches more than once, we remove the keyphrase from the golden truth when there is a match with it. That is, for each match between the predicted and actual keyphrases, we remove the keyphrase from the golden set so that it is not used to count the matches of the remaining predicted keyphrases.

We perform statistical significance tests for both the exact and partial match evaluations. As we have to compare more than two systems, we use the Randomised Tukey HSD Test~\cite{tukey,tukey_book}. These tests ensure that the family-wise error does not exceed the $\alpha$ confidence level.

In the binary evaluation task used to assess the performance of the keyphrase filtering model, we utilise a classical classification metric, accuracy. In addition, we also report the model's confusion matrices on each dataset, showing the accuracy of true positives, true negatives, false positives and false negatives.

\subsection{Experimental Settings}
Here we report the parameters used to train the supervised baselines and the proposed models. We also include the information relative to each model's inference process.

First, regarding the AKE unsupervised baselines, for \texttt{TextRank}, \texttt{SingleRank}, \texttt{TopicRank} and \texttt{YAKE}, as mentioned before, we use the \texttt{pke} toolkit to evaluate the models. We used the default parameters for all of them, extracting 15 keyphrases for each input document. On the other hand, to assess the performance of the \texttt{EmbedRank} model, we also extract 15 keyphrases per document. Additionally, we use a trade-off between informativeness and diversity of 0. This means we use the full informativeness version of the model besides diversity-aware versions, which seemed to be the best approach in previous work~\cite{gabin2022exploring}.

The last unsupervised baseline we report is \texttt{Vicuna}, using the \texttt{Vicuna 13b}\footnote{https://huggingface.co/lmsys/vicuna-13b-delta-v0} checkpoint. Then, we feed the model with the necessary guidelines to tutor the model on generating keyphrases for the input documents (\ref{appendix:vicuna}). These guidelines include how the output format should be, as well as some examples of golden keyphrases for specific documents.

In terms of the supervised baselines, first, to train and test the \texttt{KEA} model, we used the \texttt{pke} toolkit as we anticipated. We used the default parameters shown in the example script provided by the toolkit authors. Again, in line with all the methods used in AKE, we extract 15 keyphrases for each document.

Second, to train the \texttt{One2Set} model, we had to pre-process our datasets to match the model's format. In this pre-processing, we needed to have a separate validation split. We decided to use 10\% of the training data for this matter. Regarding the training parameters, we used all the default parameters provided in the one-click script provided by the model's authors except the batch size and epoch, which we adapted to our needs. We used batches of size 128 and followed an early-stopping strategy regarding training steps. We also used the default inference parameters, adjusting the batch size to 128. Finally, we found that training the model on the KPTimes dataset was impossible due to capacity errors when executing the provided script. Because of this, we do not report results on this dataset.

Regarding the \texttt{KeyBART} baseline, we follow the same approach authors follow in the paper. We only fine-tune the model on the training set of the KP20k dataset, and then we use this model to execute inference on all test datasets.

Hereafter we describe the parameters used to train the T5-based models used in this work.

For fine-tuning the AKG models, we use default parameters with a batch size of 64 per device, running on two GPUs. In the \texttt{docT5keywords vanilla} strategy, which fine-tunes the \texttt{t5-base} model on each dataset, and the \texttt{docT5keywords FlanT5} strategy, which starts from the \texttt{flan-t5-base} model, we applied an early-stopping approach. This meant training the models only until no further improvement in the F1 metric was observed on a small validation split, with a maximum limit of 1000 epochs for the smaller dataset (Inspec) and 10 epochs for the larger datasets.

When fine-tuning the model using the MAG dataset, we had to deviate from this approach due to time constraints. Instead of relying on early stopping, we trained the model for a single full epoch to cover all the documents in the dataset. Then, for the \texttt{docT5keywords MAG ft} approach, which further fine-tunes the model already trained on the MAG dataset, we used the same strategy as in the \texttt{t5-base} and \texttt{flan-t5-base} models to ensure a fair comparison between the three approaches.

As previously mentioned, we report results using two different inference approaches: greedy decoding and majority voting. For greedy decoding, we use the default parameters, setting the maximum number of tokens to be generated at 128. This configuration remains consistent across both strategies.

In the majority voting approach, we configure the model to generate 15 different outputs, applying a repetition penalty of 1.5 to prevent the model from generating the same keyphrases repeatedly. After generating the outputs, we rank the keyphrases by their frequency.

Regarding the filtering model training, we use the provided scripts by \texttt{monoT5} model authors, using all the default parameters except the batch size and the number of epochs. In terms of batch size, we used the same for all the datasets, 64 per GPU. On the other hand, the number of epochs varies depending on each dataset: 100 for Inspec, 4 for KP20k, 2 for KPTimes and 4 for KP-BioMed.

\section{Results}
\label{section:results}

\begin{table}
\centering
\caption{Present keyphrase labelling exact match evaluation results. Statistically significant improvements according to the Randomized Tukey HSD test (1M permutations, $\alpha = 0.05$) are shown with superscripted letters from $a$ to $r$.}
\label{table:present_exact_results}
\renewcommand{\arraystretch}{1.4}
\setlength{\tabcolsep}{5pt}
\footnotesize
\begin{tabular}
{l@{\hspace{1\tabcolsep}}rrrrrrrrrrrrr}
\toprule
\multicolumn{2}{l}{\multirow{2}*{Method}} & & \multicolumn{2}{c}{Inspec} & & \multicolumn{2}{c}{KP20k} & & \multicolumn{2}{c}{KPTimes} & & \multicolumn{2}{c}{KP-BioMed} \\
 & & & \multicolumn{1}{c} {F$_1$@5} & \multicolumn{1}{c} {F$_1$@$\mathcal{O}$} &
 & \multicolumn{1}{c} {F$_1$@5} & \multicolumn{1}{c} {F$_1$@$\mathcal{O}$} &
 & \multicolumn{1}{c} {F$_1$@5} & \multicolumn{1}{c} {F$_1$@$\mathcal{O}$} &
 & \multicolumn{1}{c} {F$_1$@5} & \multicolumn{1}{c} {F$_1$@$\mathcal{O}$}  \\

\cline{1-2} \cline{4-5} \cline{7-8} \cline{10-11} \cline{13-14}

\rowfont{\tiny}
& & & \multicolumn{1}{R}{$dfgh$} & \multicolumn{1}{R}{$dfgh$} & &  &  & &  & \multicolumn{1}{R}{$gh$} & &  &  \\[-5pt]
\multicolumn{2}{l}{\texttt{TextRank} ($a$)}   & & 0.293 & 0.353 & & 0.078 & 0.052 & & 0.015 & 0.011 & & 0.050 & 0.037 \\
\rowfont{\tiny}
& & & \multicolumn{1}{R}{$dfgh$} & \multicolumn{1}{R}{$cdfgh$} & & \multicolumn{1}{R}{$a$} & \multicolumn{1}{R}{$a$} & & \multicolumn{1}{R}{$ah$} & \multicolumn{1}{R}{$agh$} & & \multicolumn{1}{R}{$a$} & \multicolumn{1}{R}{$a$} \\[-5pt]
\multicolumn{2}{l}{\texttt{SingleRank} ($b$)} & & 0.301 & 0.357 & & 0.118 & 0.090 & & 0.055 & 0.041 & & 0.091 & 0.075 \\
\rowfont{\tiny}
& & & \multicolumn{1}{R}{$dfgh$} & \multicolumn{1}{R}{$dfgh$} & & \multicolumn{1}{R}{$abde$} & \multicolumn{1}{R}{$abde$} & & \multicolumn{1}{R}{$abdegh$} & \multicolumn{1}{R}{$abdegh$} & & \multicolumn{1}{R}{$abeg$} & \multicolumn{1}{R}{$abdeg$} \\[-5pt]
\multicolumn{2}{l}{\texttt{TopicRank} ($c$)}  & & 0.271 & 0.305 & & 0.145 & 0.150 & & 0.165 & 0.170 & & 0.168 & 0.168 \\
\rowfont{\tiny}
& & & \multicolumn{1}{R}{$g$} & \multicolumn{1}{R}{$g$} & & \multicolumn{1}{R}{$ab$} & \multicolumn{1}{R}{$ab$} & & \multicolumn{1}{R}{$abegh$} & \multicolumn{1}{R}{$abegh$} & & \multicolumn{1}{R}{$abeg$} & \multicolumn{1}{R}{$abeg$} \\[-5pt]
\multicolumn{2}{l}{\texttt{YAKE} ($d$)}       & & 0.196 & 0.223 & & 0.126 & 0.110 & & 0.120 & 0.104 & & 0.169 & 0.159 \\
\rowfont{\tiny}
& & & \multicolumn{1}{R}{$cdfgh$} & \multicolumn{1}{R}{$cdfgh$} & & \multicolumn{1}{R}{$abd$} & \multicolumn{1}{R}{$ab$} & & \multicolumn{1}{R}{$abgh$} & \multicolumn{1}{R}{$agh$} & & \multicolumn{1}{R}{$ab$} & \multicolumn{1}{R}{$ab$} \\[-5pt]
\multicolumn{2}{l}{\texttt{EmbedRank} ($e$)}  & & 0.324 & 0.391 & & 0.134 & 0.117 & & 0.100 & 0.085 & & 0.129 & 0.117 \\
\rowfont{\tiny}
& & &  &  & & \multicolumn{1}{R}{$abcde$} & \multicolumn{1}{R}{$abcde$} & & \multicolumn{1}{R}{$abcdegh$} & \multicolumn{1}{R}{$abcdegh$} & & \multicolumn{1}{R}{$abcdeg$} & \multicolumn{1}{R}{$abcdeg$} \\[-5pt]
\multicolumn{2}{l}{\texttt{KEA} ($f$)}        & & 0.159 & 0.186 & & 0.200 & 0.204 & & 0.214 & 0.215 & & 0.222 & 0.219 \\
\rowfont{\tiny}
& & &  &  & & \multicolumn{1}{R}{$abcdef$} & \multicolumn{1}{R}{$abcdef$} & & \multicolumn{1}{R}{$ah$} & \multicolumn{1}{R}{$h$} & & \multicolumn{1}{R}{$abe$} & \multicolumn{1}{R}{$abe$} \\[-5pt]
\multicolumn{2}{l}{\texttt{One2Set} ($g$)}    & & 0.147 & 0.149 & & 0.217 & 0.242 & & 0.053 & 0.051 & & 0.138 & 0.141 \\
\rowfont{\tiny}
& & & \multicolumn{1}{R}{$g$} &  & & \multicolumn{1}{R}{$abcdefg$} & \multicolumn{1}{R}{$abcdefg$} & &  &  & & \multicolumn{1}{R}{$abcdeg$} & \multicolumn{1}{R}{$abcdeg$} \\[-5pt]
\multicolumn{2}{l}{\texttt{KeyBART} ($h$)}    & & 0.194 & 0.196 & & 0.249 & 0.251 & & 0.017 & 0.018 & & 0.224 & 0.225 \\
\rowfont{\tiny}
& & & \multicolumn{1}{R}{$dfgh$} & \multicolumn{1}{R}{$cdfgh$} & & \multicolumn{1}{R}{$abcdefgh$} & \multicolumn{1}{R}{$abcdefgh$} & & \multicolumn{1}{R}{$abcdefgh$} & \multicolumn{1}{R}{$abcdefgh$} & & \multicolumn{1}{R}{$abcdefgh$} & \multicolumn{1}{R}{$abcdefghq$} \\[-5pt]
\multicolumn{2}{l}{\texttt{Vicuna} ($i$)}     & & 0.302 & 0.372 & & 0.269 & 0.287 & & 0.246 & 0.269 & & 0.321 & 0.336 \\

\midrule
\multicolumn{7}{l}{\texttt{docT5keywords greedy decoding}} \\
\rowfont{\tiny}
& & & \multicolumn{1}{R}{$abcdefghi$} & \multicolumn{1}{R}{$abcdefghi$} & & \multicolumn{1}{R}{$abcdefghimn$} & \multicolumn{1}{R}{$abcdefghi$} & & \multicolumn{1}{R}{$abcdefghimnopr$} & \multicolumn{1}{R}{$abcdefghimnopqr$} & & \multicolumn{1}{R}{$abcdefghiknqr$} & \multicolumn{1}{R}{$abcdefghiknpqr$} \\[-5pt]
& \texttt{vanilla} ($j$)                      & & 0.443 & 0.518 & & 0.297 & 0.303 & & 0.315 & 0.332 & & \textBF{0.343} & \textBF{0.358} \\
\rowfont{\tiny}
& & & \multicolumn{1}{R}{$abcdefghi$} & \multicolumn{1}{R}{$abcdefghi$} & & \multicolumn{1}{R}{$abcdefghi$} & \multicolumn{1}{R}{$abcdefghi$} & & \multicolumn{1}{R}{$abcdefghimnopr$} & \multicolumn{1}{R}{$abcdefghijlmnopqr$} & & \multicolumn{1}{R}{$abcdefgh$} & \multicolumn{1}{R}{$abcdefghipqr$} \\[-5pt]
& \texttt{FlanT5} ($k$)                       & & 0.450 & 0.534 & & 0.292 & 0.304 & & 0.314 & \textBF{0.345} & & 0.326 & 0.345 \\
\rowfont{\tiny}
& & & \multicolumn{1}{R}{$abcdefghi$} & \multicolumn{1}{R}{$abcdefghi$} & & \multicolumn{1}{R}{$abcdefghijkmno$} & \multicolumn{1}{R}{$abcdefghinpq$} & & \multicolumn{1}{R}{$abcdefghimno$} & \multicolumn{1}{R}{$abcdefghimnopr$} & & \multicolumn{1}{R}{$abcdefghiknqr$} & \multicolumn{1}{R}{$abcdefghiknpqr$} \\[-5pt]
& \texttt{MAG ft} ($l$)                       & & 0.443 & 0.516 & & 0.304 & \textBF{0.311} & & 0.311 & 0.327 & & 0.342 & 0.357 \\ [+5pt]

\multicolumn{7}{l}{\texttt{docT5keywords majority voting}} \\
\rowfont{\tiny}
& & & \multicolumn{1}{R}{$abcdefghi$} & \multicolumn{1}{R}{$abcdefghi$} & & \multicolumn{1}{R}{$abcdefghi$} & \multicolumn{1}{R}{$abcdefghi$} & & \multicolumn{1}{R}{$abcdefghi$} & \multicolumn{1}{R}{$abcdefghi$} & & \multicolumn{1}{R}{$abcdefghiknqr$} & \multicolumn{1}{R}{$abcdefghinpqr$} \\[-5pt]
& \texttt{vanilla} ($m$)                      & & 0.447 & 0.537 & & 0.290 & 0.303 & & 0.295 & 0.311 & & 0.342 & 0.352 \\
\rowfont{\tiny}
& & & \multicolumn{1}{R}{$abcdefghi$} & \multicolumn{1}{R}{$abcdefghi$} & & \multicolumn{1}{R}{$abcdefghi$} & \multicolumn{1}{R}{$abcdefghi$} & & \multicolumn{1}{R}{$abcdefghi$} & \multicolumn{1}{R}{$abcdefghipr$} & & \multicolumn{1}{R}{$abcdefgh$} & \multicolumn{1}{R}{$abcdefghpqr$} \\[-5pt]
& \texttt{FlanT5} ($n$)                       & & 0.450 & 0.542 & & 0.286 & 0.300 & & 0.296 & 0.315 & & 0.328 & 0.339 \\
\rowfont{\tiny}
& & & \multicolumn{1}{R}{$abcdefghi$} & \multicolumn{1}{R}{$abcdefghi$} & & \multicolumn{1}{R}{$abcdefghimn$} & \multicolumn{1}{R}{$abcdefghinpq$} & & \multicolumn{1}{R}{$abcdefghi$} & \multicolumn{1}{R}{$abcdefghi$} & & \multicolumn{1}{R}{$abcdefghiknqr$} & \multicolumn{1}{R}{$abcdefghinpqr$} \\[-5pt]
& \texttt{MAG ft} ($o$)                       & & 0.450 & 0.544 & & 0.296 & 0.310 & & 0.293 & 0.307 & & 0.342 & 0.351 \\ [+5pt]

\multicolumn{7}{l}{\texttt{docT5keywords greedy decoding + keyFilT5r}} \\
\rowfont{\tiny}
& & & \multicolumn{1}{R}{$abcdefghi$} & \multicolumn{1}{R}{$abcdefghi$} & & \multicolumn{1}{R}{$abcdefghijkmno$} & \multicolumn{1}{R}{$abcdefghi$} & & \multicolumn{1}{R}{$abcdefghimno$} & \multicolumn{1}{R}{$abcdefghino$} & & \multicolumn{1}{R}{$abcdefghiknq$} & \multicolumn{1}{R}{$abcdefghq$} \\[-5pt]
& \texttt{vanilla} ($p$)                    & & 0.472 & 0.537 & & 0.304 & 0.300 & & 0.305 & 0.304 & & 0.336 & 0.331 \\
\rowfont{\tiny}
& & & \multicolumn{1}{R}{$abcdefghi$} & \multicolumn{1}{R}{$abcdefghi$} & & \multicolumn{1}{R}{$abcdefghijkmno$} & \multicolumn{1}{R}{$abcdefghi$} & & \multicolumn{1}{R}{$abcdefghimnopr$} & \multicolumn{1}{R}{$abcdefghimopr$} & & \multicolumn{1}{R}{$abcdefgh$} & \multicolumn{1}{R}{$abcdefgh$} \\[-5pt]
& \texttt{FlanT5} ($q$)                     & & \textBF{0.479} & \textBF{0.552} & & 0.304 & 0.301 & & \textBF{0.316} & 0.320 & & 0.325 & 0.321 \\
\rowfont{\tiny}
& & & \multicolumn{1}{R}{$abcdefghi$} & \multicolumn{1}{R}{$abcdefghi$} & & \multicolumn{1}{R}{$abcdefghijkmno$} & \multicolumn{1}{R}{$abcdefghi$} & & \multicolumn{1}{R}{$abcdefghimno$} & \multicolumn{1}{R}{$abcdefghi$} & & \multicolumn{1}{R}{$abcdefghiq$} & \multicolumn{1}{R}{$abcdefgh$} \\[-5pt]
& \texttt{MAG ft} ($r$)                     & & 0.472 & 0.532 & & \textBF{0.309} & 0.306 & & 0.306 & 0.304 & & 0.333 & 0.329 \\

\bottomrule
\end{tabular}
\end{table}

This section presents the performance of the different \texttt{docT5keywords} variants compared to several AKE and AKG techniques.
We structure the analysis across four distinct experiments.
The first experiment (\S~\ref{sec:e1}) reports results from the exact match evaluation for present keyphrases.
The second experiment (\S~\ref{sec:e2}) assesses the models' ability to accurately predict absent keyphrases.
In the third experiment (\S~\ref{sec:e3}), we explore the impact of keyphrase sorting during the fine-tuning phase.
Lastly, the fourth experiment (\S~\ref{sec:e4}) evaluates the effectiveness of our filtering model through a binary assessment of whether a keyphrase is relevant to a document. 
Additionally, partial match evaluation results can be found in Appendix~\ref{appendix:partial}.

\subsection{Experiment 1: Exact Match Evaluation on Present Keyphrases} \label{sec:e1}
In this experiment, we detail the results of all our fine-tuning variants of the \texttt{docT5keywords} model as explained in Section~\ref{sec:proposal}. The first one (\texttt{vanilla}) shows the results of fine-tuning a \texttt{t5-base} model on the training split of each dataset for the AKG task. The second one, \texttt{FlanT5} follows the same approach as the first one using \texttt{FlanT5} as the base model. The third one, \texttt{MAG ft}, refers to the approach that performs a large fine-tuning on a large dataset (MAG) and then continues the fine-tuning process in the training split of each individual dataset.

Additionally, we provide results for both the majority voting inference approach and the keyphrase filtering method, highlighting their impact on the performance of each fine-tuning strategy.

Table~\ref{table:present_exact_results} reports the results of the different methods on four keyphrase extraction/generation datasets using the exact match evaluation technique (\S~\ref{subsec:em}) over the present keyphrases (\S~\ref{subsec:pkps}). Exact match is the more challenging evaluation scenario because it requires identifying whole golden keyphrases.

First, the traditional methods, such as \texttt{TextRank}, \texttt{SingleRank}, and \texttt{TopicRank}, generally exhibit lower performance in this evaluation. Even the relatively better-performing baseline, \texttt{EmbedRank}, struggles to achieve competitive results. These results underscore the limitations of traditional unsupervised keyphrase extraction techniques, particularly when applied to large and diverse datasets.

In contrast, the \texttt{docT5keywords} model and its various fine-tuning strategies demonstrate significant improvements in keyphrase extraction accuracy. The variants of this model—\texttt{vanilla}, \texttt{FlanT5}, and \texttt{MAG ft}—consistently outperform the baseline methods across all datasets, with particularly high differences observed in the Inspec and KP-BioMed datasets. The \texttt{FlanT5} variant, stands out as the top performer, achieving the highest F$_1$ scores in several evaluations.

Further enhancing the results of the base models, we explore the impact of majority voting and keyphrase filtering techniques. The integration of these methods with the \texttt{docT5keywords} model yields slight yet consistent improvements. The majority voting strategy, particularly in the \texttt{vanilla} and \texttt{FlanT5} variants, shows an advantage in datasets like Inspec and KP-BioMed, where it slightly outperforms the basic fine-tuning methods. Similarly, the keyphrase filtering approach (\texttt{keyFilT5r}) matches or even surpasses the best models in certain cases, particularly when combined with the \texttt{FlanT5} variant. This combination achieves the highest F$_1$ scores on datasets such as Inspec and KPTimes, further solidifying the effectiveness of these techniques.

The statistical analysis accompanying these results, using the Randomised Tukey HSD test, underscores the robustness of the improvements observed with the \texttt{docT5keywords} models.

\begin{tcolorbox}[left*=3mm, right*=3mm, arc=1mm, boxrule=0pt, colback=BoxColor, colframe=white, title=Experiment 1 highlights, colbacktitle=BoxColorTitle, toptitle=1mm, bottomtitle=1mm]
    The \texttt{docT5keywords} model significantly outperforms baseline methods in keyphrase extraction. Incorporating advanced inference techniques like majority voting further enhances performance, while the filtering post-processing step also yields impressive results. Overall, \texttt{docT5keywords} demonstrates superior effectiveness in extracting keyphrases across various datasets.
\end{tcolorbox}

\subsection{Experiment 2: Exact Match Evaluation on Absent Keyphrases} \label{sec:e2}
\begin{table}
\centering
\caption{Absent keyphrase labelling exact match evaluation results. Statistically significant improvements according to the Randomized Tukey HSD test (1M permutations, $\alpha = 0.05$) are shown with superscripted letters from $a$ to $r$.}
\label{table:absent_exact_results}
\renewcommand{\arraystretch}{1.4}
\setlength{\tabcolsep}{5pt}
\footnotesize
\begin{tabular}
{l@{\hspace{1\tabcolsep}}rrrrrrrrrrrrr}
\toprule
\multicolumn{2}{l}{\multirow{2}*{Method}} & & \multicolumn{2}{c}{Inspec} & & \multicolumn{2}{c}{KP20k} & & \multicolumn{2}{c}{KPTimes} & & \multicolumn{2}{c}{KP-BioMed} \\
 & & & \multicolumn{1}{c} {F$_1$@5} & \multicolumn{1}{c} {F$_1$@$\mathcal{O}$} &
 & \multicolumn{1}{c} {F$_1$@5} & \multicolumn{1}{c} {F$_1$@$\mathcal{O}$} &
 & \multicolumn{1}{c} {F$_1$@5} & \multicolumn{1}{c} {F$_1$@$\mathcal{O}$} &
 & \multicolumn{1}{c} {F$_1$@5} & \multicolumn{1}{c} {F$_1$@$\mathcal{O}$}  \\

\cline{1-2} \cline{4-5} \cline{7-8} \cline{10-11} \cline{13-14}

\rowfont{\scriptsize}
& & &  &  & & \multicolumn{1}{R}{$ijklmnopqr$} & \multicolumn{1}{R}{$jklmnopqr$} & & \multicolumn{1}{R}{$h$} &  & &  &  \\[-5pt]
\multicolumn{2}{l}{\texttt{One2Set} ($g$)}    & & 0.007 & 0.006 & & 0.029 & 0.023 & & 0.006 & 0.005 & & 0.008 & 0.006 \\
\rowfont{\scriptsize}
& & &  &  & & \multicolumn{1}{R}{$ijklmnopqr$} & \multicolumn{1}{R}{$gijklmnopqr$} & &  &  & & \multicolumn{1}{R}{$gpqr$} & \multicolumn{1}{R}{$gpqr$} \\[-5pt]
\multicolumn{2}{l}{\texttt{KeyBART} ($h$)}    & & 0.012 & 0.011 & & \textBF{0.030} & \textBF{0.028} & & 0.002 & 0.002 & & 0.012 & 0.009 \\
\rowfont{\scriptsize}
& & &  &  & & \multicolumn{1}{R}{$jklmnopqr$} & \multicolumn{1}{R}{$jklmnopqr$} & & \multicolumn{1}{R}{$gh$} & \multicolumn{1}{R}{$gh$} & & \multicolumn{1}{R}{$ghjklmnopqr$} & \multicolumn{1}{R}{$ghjklmnopqr$} \\[-5pt]
\multicolumn{2}{l}{\texttt{Vicuna} ($i$)}     & & 0.017 & 0.008 & & 0.025 & 0.024 & & 0.025 & 0.022 & & \textBF{0.018} & \textBF{0.014} \\

\midrule
\multicolumn{7}{l}{\texttt{docT5keywords greedy decoding}} \\
\rowfont{\scriptsize}
& & & \multicolumn{1}{R}{$gh$} &  & & \multicolumn{1}{R}{$pqr$} & \multicolumn{1}{R}{$pqr$} & & \multicolumn{1}{R}{$ghipqr$} & \multicolumn{1}{R}{$ghipqr$} & & \multicolumn{1}{R}{$ghknpqr$} & \multicolumn{1}{R}{$gpqr$} \\[-5pt]
& \texttt{vanilla} ($j$)                      & & 0.027 & 0.017 & & 0.019 & 0.017 & & 0.088 & 0.086 & & 0.015 & 0.010 \\
\rowfont{\scriptsize}
& & & \multicolumn{1}{R}{$gh$} & \multicolumn{1}{R}{$g$} & & \multicolumn{1}{R}{$pqr$} & \multicolumn{1}{R}{$pqr$} & & \multicolumn{1}{R}{$ghijmpqr$} & \multicolumn{1}{R}{$ghijmpqr$} & & \multicolumn{1}{R}{$gpqr$} & \multicolumn{1}{R}{$gpqr$} \\[-5pt]
& \texttt{FlanT5} ($k$)                       & & 0.028 & 0.023 & & 0.020 & 0.018 & & 0.102 & 0.100 & & 0.013 & 0.009 \\
\rowfont{\scriptsize}
& & & \multicolumn{1}{R}{$ghijkmnpqr$} & \multicolumn{1}{R}{$ghijmpq$} & & \multicolumn{1}{R}{$jpqr$} & \multicolumn{1}{R}{$jmpqr$} & & \multicolumn{1}{R}{$ghijmpqr$} & \multicolumn{1}{R}{$ghijmpqr$} & & \multicolumn{1}{R}{$ghknpqr$} & \multicolumn{1}{R}{$gpqr$} \\[-5pt]
& \texttt{MAG ft} ($l$)                       & & \textBF{0.043} & \textBF{0.039} & & 0.022 & 0.019 & & 0.099 & 0.096 & & 0.014 & 0.010 \\ [+5pt]

\multicolumn{7}{l}{\texttt{docT5keywords majority voting}} \\
\rowfont{\scriptsize}
& & & \multicolumn{1}{R}{$gh$} &  & & \multicolumn{1}{R}{$pqr$} & \multicolumn{1}{R}{$pqr$} & & \multicolumn{1}{R}{$ghipqr$} & \multicolumn{1}{R}{$ghijpqr$} & & \multicolumn{1}{R}{$ghknpqr$} & \multicolumn{1}{R}{$gpqr$} \\[-5pt]
& \texttt{vanilla} ($m$)                      & & 0.028 & 0.018 & & 0.020 & 0.017 & & 0.091 & 0.091 & & 0.015 & 0.010 \\
\rowfont{\scriptsize}
& & & \multicolumn{1}{R}{$gh$} & \multicolumn{1}{R}{$gi$} & & \multicolumn{1}{R}{$jpqr$} & \multicolumn{1}{R}{$pqr$} & & \multicolumn{1}{R}{$ghijpqr$} & \multicolumn{1}{R}{$ghijklmopqr$} & & \multicolumn{1}{R}{$gpqr$} & \multicolumn{1}{R}{$gpqr$} \\[-5pt]
& \texttt{FlanT5} ($n$)                       & & 0.027 & 0.025 & & 0.022 & 0.018 & & \textBF{0.107} & \textBF{0.107} & & 0.013 & 0.010 \\
\rowfont{\scriptsize}
& & & \multicolumn{1}{R}{$ghipq$} & \multicolumn{1}{R}{$ghip$} & & \multicolumn{1}{R}{$jkmpqr$} & \multicolumn{1}{R}{$jmpqr$} & & \multicolumn{1}{R}{$ghijmpqr$} & \multicolumn{1}{R}{$ghijlmpqr$} & & \multicolumn{1}{R}{$ghpqr$} & \multicolumn{1}{R}{$gpqr$} \\[-5pt]
& \texttt{MAG ft} ($o$)                       & & 0.039 & 0.031 & & 0.022 & 0.019 & & 0.101 & 0.102 & & 0.014 & 0.010 \\ [+5pt]

\multicolumn{7}{l}{\texttt{docT5keywords greedy decoding + keyFilT5r}} \\
\rowfont{\scriptsize}
& & &  &  & &  &  & & \multicolumn{1}{R}{$ghi$} & \multicolumn{1}{R}{$ghi$} & &  &  \\[-5pt]
& \texttt{vanilla} ($p$)                    & & 0.016 & 0.013 & & 0.012 & 0.011 & & 0.062 & 0.060 & & 0.008 & 0.007 \\
\rowfont{\scriptsize}
& & &  &  & &  &  & & \multicolumn{1}{R}{$ghipr$} & \multicolumn{1}{R}{$ghipr$} & &  &  \\[-5pt]
& \texttt{FlanT5} ($q$)                     & & 0.018 & 0.015 & & 0.012 & 0.012 & & 0.077 & 0.074 & & 0.007 & 0.006 \\
\rowfont{\scriptsize}
& & & \multicolumn{1}{R}{$gh$} & \multicolumn{1}{R}{$g$} & &  &  & & \multicolumn{1}{R}{$ghip$} & \multicolumn{1}{R}{$ghip$} & &  &  \\[-5pt]
& \texttt{MAG ft} ($r$)                     & & 0.026 & 0.023 & & 0.013 & 0.012 & & 0.069 & 0.068 & & 0.008 & 0.007 \\

\bottomrule
\end{tabular}
\end{table}

Following the previous experiment that focused on present keyphrases using exact match evaluation, we now turn our attention to the more challenging task of absent keyphrase labelling (\S~\ref{subsec:akps}). This task is particularly difficult because it requires models to go beyond surface-level information and generate keyphrases that capture the underlying meaning or themes of the document. These results are reported in Table~\ref{table:absent_exact_results}.

In general, the baseline models--\texttt{One2Set}, \texttt{KeyBART}, and \texttt{Vicuna}--show limited effectiveness in accurately labelling absent keyphrases. For example, \texttt{One2Set} consistently yields low F$_1$ scores across all datasets, with the highest being 0.029 on KP20k. This poor performance underscores the challenges faced by traditional approaches in capturing absent keyphrases, which often require a deeper understanding of context and semantics beyond what these models can provide.

\texttt{KeyBART} and \texttt{Vicuna}, while slightly better, still struggle in some cases. For example, \texttt{KeyBART} achieves a marginal improvement on KP20k, but this is still far from ideal, especially when compared to more sophisticated methods. Similarly, \texttt{Vicuna} performs slightly better on the KP20k and KP-BioMed datasets, but this improvement is inconsistent across other datasets.

Regarding the \texttt{docT5keywords} model, \texttt{MAG ft} stands out, particularly on the Inspec dataset. This represents a substantial improvement over baseline methods.  The performance of \texttt{MAG ft} is also remarkable on other datasets such as KPTimes where the baseline methods obtain significantly lower scores.

Further analysis reveals that combining the \texttt{docT5keywords} model with advanced inference strategies such as majority voting and keyphrase filtering can yield additional performance gains. For instance, the \texttt{FlanT5} variant, when combined with majority voting, achieves the highest scores on the KPTimes dataset.

On the other hand, keyphrase filtering techniques, tend to yield worse results than variants without the filtering model. This means that the proposed filtering method tends to filter out some relevant absent keyphrases. Identifying absent keyphrases as relevant or not is the most difficult task these models face, as it requires understanding implicit information and deciding what should be included, even when it is not explicitly mentioned in the text. The performance drop in predicting the relevance of absent keyphrases can be attributed to the imbalance in the classifier's training data, where the number of absent keyphrases is much smaller than the present ones. This imbalance makes it more challenging for the model to generalise patterns related to absent keyphrases, leading to reduced performance.

The statistical significance of these results is once again validated through the Randomised Tukey HSD test, which confirms that the improvements by \texttt{docT5keywords} and its variants are not just marginal but statistically significant.

In conclusion, While traditional methods struggle absent keyphrases, the \texttt{docT5keywords} model and its variants, particularly when combined with techniques like majority voting and keyphrase filtering, offer a substantial improvement.

\begin{tcolorbox}[left*=3mm, right*=3mm, arc=1mm, boxrule=0pt, colback=BoxColor, colframe=white, title=Experiment 2 highlights, colbacktitle=BoxColorTitle, toptitle=1mm, bottomtitle=1mm]
The \texttt{docT5keywords} model significantly outperforms baselines in two out of four datasets and achieves comparable results in the remaining two. The majority voting technique proves effective, while the filtering approach falls short due to the complexity of accurately labelling absent keyphrases as relevant or not.
\end{tcolorbox}

\subsection{Experiment 3: Fine-tuning with Sorted Keyphrases} \label{sec:e3}
We conduct an additional experiment to investigate the impact of keyphrase order on the training of our keyphrase generation models. Previous studies~\cite{meng2021empirical} have suggested that the order of keyphrases can significantly affect the performance of such models. To validate those findings, we explore the influence of keyphrase order during the fine-tuning process. 

Contrary to the findings of previous studies, our experimental results (Table~\ref{table:sorted}) reveal a surprising outcome regarding the impact of keyphrase order on the performance of our keyphrase generation models. Specifically, we discovered that ordering the keyphrases according to the convention of presenting present keyphrases before absent keyphrases did not improve model performance. Our findings indicate that adhering to this ordering scheme even resulted in a deterioration of the model performance. These unexpected results challenge the existing assumptions. 

\begin{table}
\centering
\caption{Non-sorted vs sorted keyphrases training strategies. We report results using the exact match evaluation approach without differentiating between present and absent keyphrases. Statistically significant improvements according to the Randomized Tukey HSD test (1M permutations, $\alpha = 0.05$) are shown with superscripted letters $j$ and $s$ respectively.}
\label{table:sorted}
\renewcommand{\arraystretch}{1.7}
\setlength{\tabcolsep}{7pt}
\footnotesize
\begin{tabular}
{l@{\hspace{1\tabcolsep}}rrrrrrrrrrrrr}
\toprule
\multicolumn{2}{l}{\multirow{2}*{Method}} & & \multicolumn{2}{c}{Inspec} & & \multicolumn{2}{c}{KP20k} & & \multicolumn{2}{c}{KPTimes} & & \multicolumn{2}{c}{KP-BioMed} \\
& & & \multicolumn{1}{c} {F$_1$@5} & \multicolumn{1}{c} {F$_1$@$\mathcal{O}$} &
 & \multicolumn{1}{c} {F$_1$@5} & \multicolumn{1}{c} {F$_1$@$\mathcal{O}$} &
 & \multicolumn{1}{c} {F$_1$@5} & \multicolumn{1}{c} {F$_1$@$\mathcal{O}$} &
 & \multicolumn{1}{c} {F$_1$@5} & \multicolumn{1}{c} {F$_1$@$\mathcal{O}$}  \\

\cline{1-2} \cline{4-5} \cline{7-8} \cline{10-11} \cline{13-14}
\multicolumn{2}{l}{\texttt{docT5keywords greedy decoding}} \\[-7pt]
\rowfont{\scriptsize}
& & & \multicolumn{1}{R}{$s$} & \multicolumn{1}{R}{$s$} & & \multicolumn{1}{R}{$s$} & \multicolumn{1}{R}{$s$} & & \multicolumn{1}{R}{$s$} & \multicolumn{1}{R}{$s$} & & \multicolumn{1}{R}{$s$} & \multicolumn{1}{R}{$s$} \\[-7pt]
& \texttt{vanilla} ($j$)                     & & \textBF{0.403} & \textBF{0.488} & & \textBF{0.248} & \textBF{0.249} & & \textBF{0.317} & \textBF{0.321} & & \textBF{0.302} & \textBF{0.308} \\
& \texttt{pres-abs sorted} ($s$)             & & 0.379 & 0.451 & & 0.239 & 0.243 & & 0.305 & 0.308 & & 0.290 & 0.295 \\

\bottomrule
\end{tabular}
\end{table}

\begin{tcolorbox}[left*=3mm, right*=3mm, arc=1mm, boxrule=0pt, colback=BoxColor, colframe=white, title=Experiment 3 highlights, colbacktitle=BoxColorTitle, toptitle=1mm, bottomtitle=1mm]
Our experiment found that ordering keyphrases with present ones before absent ones actually worsens model performance, contradicting previous studies suggesting keyphrase order is crucial. This challenges existing assumptions about the impact of keyphrase ordering.
\end{tcolorbox}

\subsection{Experiment 4: Keyphrase Filtering Binary Evaluation} \label{sec:e4}
\begin{figure}
\captionsetup[subfigure]{labelformat=empty}
\setlength{\tabcolsep}{2pt}
\centering
\begin{tabular}{cccc}
\begin{subfigure}[t]{0.23\columnwidth} \centering \includegraphics[height=\columnwidth]{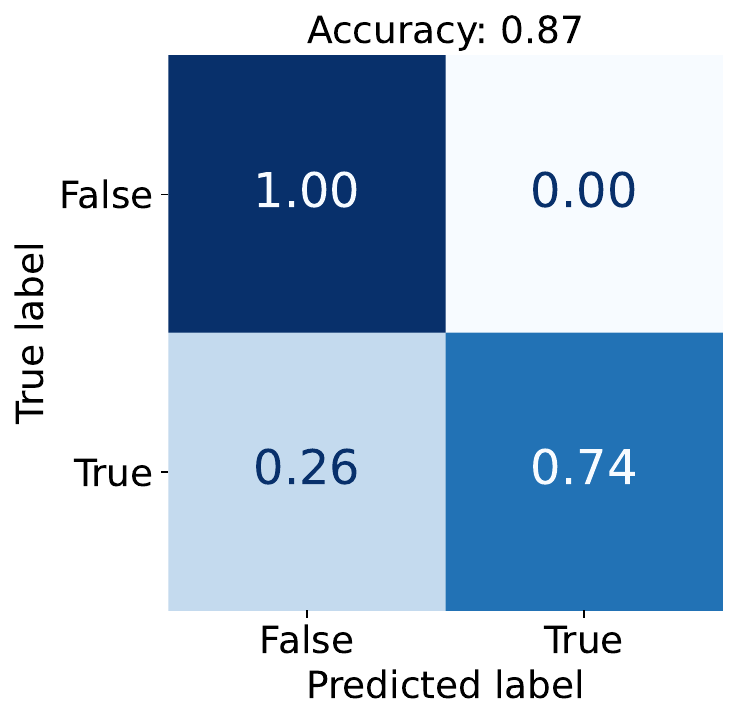} \caption{Inspec} 
\end{subfigure} &
\begin{subfigure}[t]{0.23\columnwidth} \centering \includegraphics[height=\columnwidth]{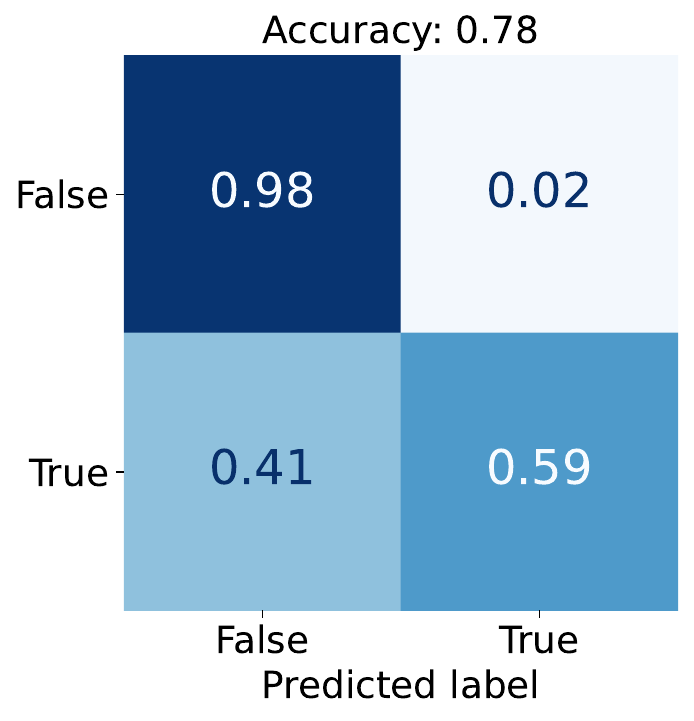} \caption{KP20k} 
\end{subfigure} &
\begin{subfigure}[t]{0.23\columnwidth} \centering \includegraphics[height=\columnwidth]{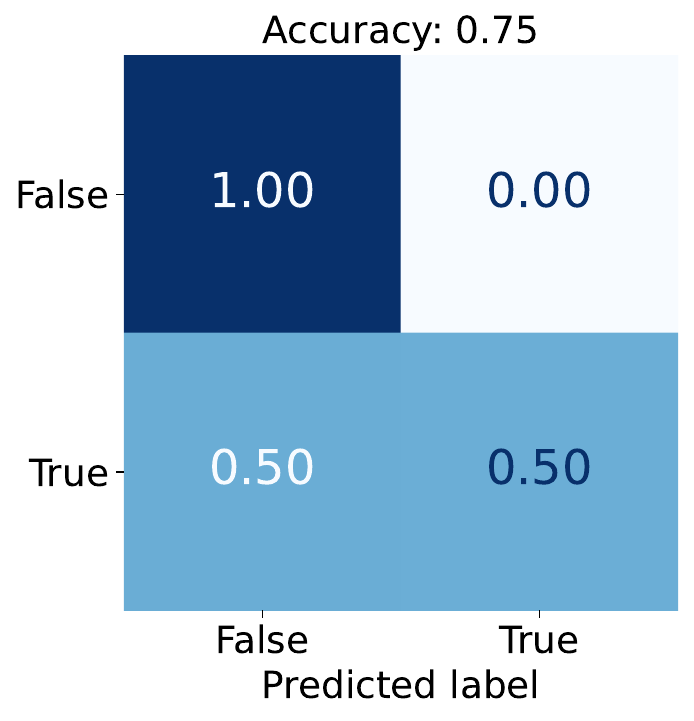} \caption{KPTimes} 
\end{subfigure} &
\begin{subfigure}[t]{0.23\columnwidth} \centering \includegraphics[height=\columnwidth]{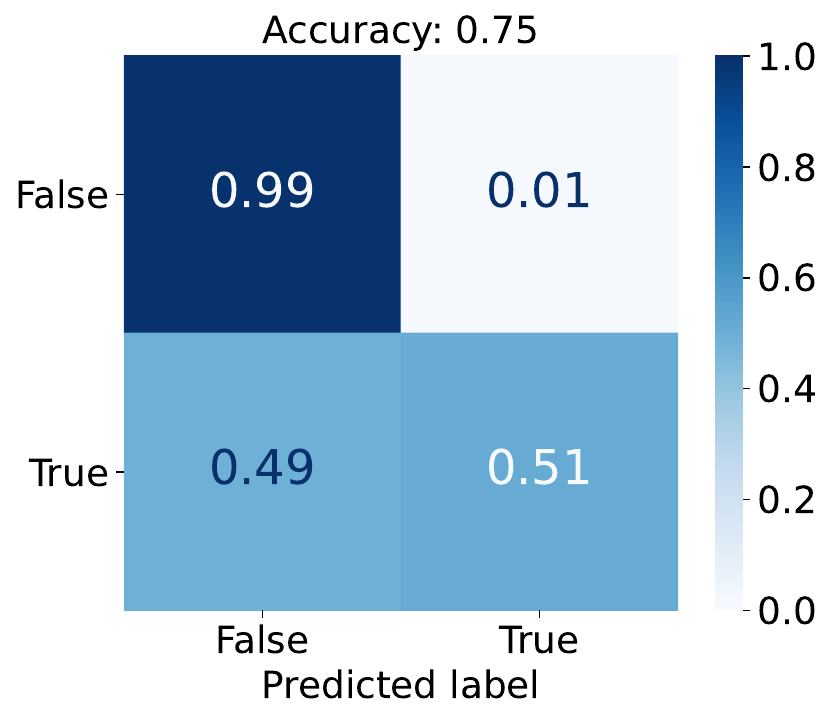} \caption{KP-BioMed} 
\end{subfigure}
\end{tabular}
\caption{\texttt{keyFilT5r} binary evaluation accuracy confusion matrices.}
\label{fig:confusion_matrices}
\end{figure}

Wrapping up the results section, Figure~\ref{fig:confusion_matrices} illustrates the binary evaluation (\S~\ref{subsec:filtering_eval}) accuracy confusion matrices for the filtering model across the four evaluation datasets: Inspec, KP20k, KPTimes, and KP-BioMed. The accuracy metrics for each dataset are presented at the top of each confusion matrix, reflecting how well the model distinguishes between true and false keyphrases.

The confusion matrix for the Inspec dataset shows the highest accuracy at 0.87. Here, the model is highly effective at predicting false keyphrases. However, there is still a 26\% false negative rate, where true keyphrases are mistakenly labelled as false.

For the KP20k dataset, the model's accuracy drops to 0.78. The matrix reveals that while the model maintains a high accuracy for false keyphrases at 0.98, its accuracy for true keyphrases falls to 0.59, suggesting a noticeable increase in false negatives. This indicates that the model struggles more with identifying true keyphrases correctly in this dataset.

The performance on the KPTimes and KP-BioMed datasets is similar, both achieving an accuracy of 0.75. In both cases, the model perfectly predicts false keyphrases but only achieves around 50\% accuracy for true keyphrases. This implies a balance between true positives and false negatives, highlighting that the model is as likely to miss true keyphrases as it is to identify them correctly.

Overall, the results suggest that the \texttt{keyFilT5r} model effectively filters out false keyphrases, achieving near-perfect accuracy across all datasets. This ability to consistently reject incorrect keyphrases while still identifying a portion of the true keyphrases is valuable in real-world applications. In scenarios where trust and reliability are crucial, such as in professional or academic settings, minimising the risk of generative models producing hallucinated or irrelevant keyphrases is essential. 

\begin{tcolorbox}[left*=3mm, right*=3mm, arc=1mm, boxrule=0pt, colback=BoxColor, colframe=white, title=Experiment 4 highlights, colbacktitle=BoxColorTitle, toptitle=1mm, bottomtitle=1mm]
The \texttt{keyFilT5r} model excels at filtering out false positives, achieving near-perfect accuracy in rejecting incorrect keyphrases, such as 0.98 on KP20k. However, it faces difficulties in accurately identifying true keyphrases, with accuracies ranging from 0.59 to 0.50 across different datasets. This suggests the model is effective at minimising false positives but struggles with true keyphrase detection.
\end{tcolorbox}

\section{Conclusions}
\label{sec:conclusions}

In this study, we conducted an in-depth evaluation of several keyphrase generation and filtering strategies. The results from exact and partial match evaluations were mainly aligned, showing that models tend to perform consistently across different datasets and evaluation types. The partial match results, as expected, yielded higher scores due to the less strict criteria, underscoring the trade-off between strict accuracy and practical flexibility in real-world applications. This suggests that the models are proficient at capturing keyphrase meanings, even if they don't match the exact wording.

In a real-world context, the ability to prevent generative models from producing hallucinations—untrue or irrelevant keyphrases—becomes vital to gain user trust. The methods evaluated in this study, particularly those using filtering techniques, showed promise in achieving this balance. A model that consistently filters out false keyphrases while successfully identifying relevant ones ensures that users can trust the output. This is essential in fields such as content generation, academic research, or any domain where the reliability of machine-generated information is critical for decision-making.

\newpage
\appendix
\section{Partial Match Evaluation}
\label{appendix:partial}
In this appendix, we present the results of the partial match (\S~\ref{subsec:pm}) evaluation for both present and absent keyphrases. Results of this evaluation are show in Tables~\ref{table:present_partial_results} and \ref{table:ansent_partial_results}. As expected, the findings from this evaluation align closely with those observed in the exact match evaluation. The methods that demonstrated strong performance in the exact match scenario similarly excelled in the partial match evaluation, indicating a consistent ability to identify key concepts across different levels of matching criteria.

It is important to note that the scores obtained in the partial match evaluation are generally higher than those in the exact match evaluation. This increase in scores can be attributed to the reduced difficulty of the partial match task, which allows for more flexibility in identifying keyphrases. Partial matching relaxes the strict requirement for an exact word-for-word correspondence, instead rewarding models for capturing the essential meaning or components of a keyphrase. Consequently, this evaluation provides a broader measure of a model's capability to generate relevant keyphrases, further reinforcing the relative strengths and weaknesses observed in the exact match results.

\begin{table}
\centering
\caption{Absent keyphrase labelling partial match evaluation results. Statistically significant improvements according to the Randomized Tukey HSD test (1M permutations, $\alpha = 0.05$) are shown with superscripted letters from $a$ to $r$.}
\label{table:ansent_partial_results}
\renewcommand{\arraystretch}{1.25}
\setlength{\tabcolsep}{5pt}
\footnotesize
\begin{tabular}
{l@{\hspace{1\tabcolsep}}rrrrrrrrrrrrr}
\toprule
\multicolumn{2}{l}{\multirow{2}*{Method}} & & \multicolumn{2}{c}{Inspec} & & \multicolumn{2}{c}{KP20k} & & \multicolumn{2}{c}{KPTimes} & & \multicolumn{2}{c}{KP-BioMed} \\
 & & & \multicolumn{1}{c} {F$_1$@5} & \multicolumn{1}{c} {F$_1$@$\mathcal{O}$} &
 & \multicolumn{1}{c} {F$_1$@5} & \multicolumn{1}{c} {F$_1$@$\mathcal{O}$} &
 & \multicolumn{1}{c} {F$_1$@5} & \multicolumn{1}{c} {F$_1$@$\mathcal{O}$} &
 & \multicolumn{1}{c} {F$_1$@5} & \multicolumn{1}{c} {F$_1$@$\mathcal{O}$}  \\

\cline{1-2} \cline{4-5} \cline{7-8} \cline{10-11} \cline{13-14}

\rowfont{\scriptsize}
& & &  & \multicolumn{1}{R}{$h$} & & \multicolumn{1}{R}{$hkmno$} & \multicolumn{1}{R}{$hmnpqr$} & & \multicolumn{1}{R}{$h$} & \multicolumn{1}{R}{$h$} & & \multicolumn{1}{R}{$hn$} & \multicolumn{1}{R}{$hpqr$} \\[-5pt]
\multicolumn{2}{l}{\texttt{One2Set} ($g$)}    & & 0.179 & 0.334 & & 0.146 & 0.213 & & 0.080 & 0.159 & & 0.147 & 0.266 \\
\rowfont{\scriptsize}
& & &  &  & &  &  & &  &  & &  &  \\[-5pt]
\multicolumn{2}{l}{\texttt{KeyBART} ($h$)}    & & 0.159 & 0.243 & & 0.123 & 0.177 & & 0.018 & 0.024 & & 0.109 & 0.183 \\
\rowfont{\scriptsize}
& & & \multicolumn{1}{R}{$ghmnopqr$} & \multicolumn{1}{R}{$hpr$} & & \multicolumn{1}{R}{$hkmno$} & \multicolumn{1}{R}{$ghjklmnopqr$} & & \multicolumn{1}{R}{$gh$} & \multicolumn{1}{R}{$gh$} & & \multicolumn{1}{R}{$ghklmno$} & \multicolumn{1}{R}{$ghjklmnopqr$} \\[-5pt]
\multicolumn{2}{l}{\texttt{Vicuna} ($i$)}     & & \textBF{0.213} & \textBF{0.363} & & 0.150 & \textBF{0.227} & & 0.149 & 0.235 & & 0.154 & \textBF{0.279} \\

\midrule
\multicolumn{7}{l}{\texttt{docT5keywords greedy decoding}} \\
\rowfont{\scriptsize}
& & & \multicolumn{1}{R}{$h$} & \multicolumn{1}{R}{$h$} & & \multicolumn{1}{R}{$hkmno$} & \multicolumn{1}{R}{$hpqr$} & & \multicolumn{1}{R}{$ghimno$} & \multicolumn{1}{R}{$ghipqr$} & & \multicolumn{1}{R}{$hkmno$} & \multicolumn{1}{R}{$ghpqr$} \\[-5pt]
& \texttt{vanilla} ($j$)                      & & 0.190 & 0.337 & & 0.147 & 0.210 & & 0.232 & 0.301 & & 0.150 & 0.271 \\
\rowfont{\scriptsize}
& & & \multicolumn{1}{R}{$h$} & \multicolumn{1}{R}{$h$} & & \multicolumn{1}{R}{$hmno$} & \multicolumn{1}{R}{$hpqr$} & & \multicolumn{1}{R}{$ghimo$} & \multicolumn{1}{R}{$ghijmpqr$} & & \multicolumn{1}{R}{$h$} & \multicolumn{1}{R}{$hpqr$} \\[-5pt]
& \texttt{FlanT5} ($k$)                       & & 0.190 & 0.346 & & 0.142 & 0.209 & & 0.226 & 0.312 & & 0.143 & 0.268 \\
\rowfont{\scriptsize}
& & & \multicolumn{1}{R}{$h$} & \multicolumn{1}{R}{$h$} & & \multicolumn{1}{R}{$hkmno$} & \multicolumn{1}{R}{$hpqr$} & & \multicolumn{1}{R}{$ghikmno$} & \multicolumn{1}{R}{$ghijmpqr$} & & \multicolumn{1}{R}{$hkn$} & \multicolumn{1}{R}{$hpqr$} \\[-5pt]
& \texttt{MAG ft} ($l$)                       & & 0.197 & 0.342 & & 0.148 & 0.210 & & 0.235 & 0.309 & & 0.149 & 0.271 \\ [+5pt]

\multicolumn{7}{l}{\texttt{docT5keywords majority voting}} \\
\rowfont{\scriptsize}
& & & \multicolumn{1}{R}{$h$} & \multicolumn{1}{R}{$h$} & & \multicolumn{1}{R}{$h$} & \multicolumn{1}{R}{$hpqr$} & & \multicolumn{1}{R}{$ghi$} & \multicolumn{1}{R}{$ghipqr$} & & \multicolumn{1}{R}{$hn$} & \multicolumn{1}{R}{$hpqr$} \\[-5pt]
& \texttt{vanilla} ($m$)                      & & 0.186 & 0.340 & & 0.133 & 0.208 & & 0.209 & 0.302 & & 0.145 & 0.270 \\
\rowfont{\scriptsize}
& & & \multicolumn{1}{R}{$h$} & \multicolumn{1}{R}{$h$} & & \multicolumn{1}{R}{$h$} & \multicolumn{1}{R}{$hpqr$} & & \multicolumn{1}{R}{$ghim$} & \multicolumn{1}{R}{$ghijmpqr$} & & \multicolumn{1}{R}{$h$} & \multicolumn{1}{R}{$hpqr$} \\[-5pt]
& \texttt{FlanT5} ($n$)                       & & 0.181 & 0.345 & & 0.134 & 0.207 & & 0.221 & \textBF{0.314} & & 0.141 & 0.267 \\
\rowfont{\scriptsize}
& & & \multicolumn{1}{R}{$h$} & \multicolumn{1}{R}{$h$} & & \multicolumn{1}{R}{$h$} & \multicolumn{1}{R}{$hpqr$} & & \multicolumn{1}{R}{$ghim$} & \multicolumn{1}{R}{$ghijmpqr$} & & \multicolumn{1}{R}{$hn$} & \multicolumn{1}{R}{$hpqr$} \\[-5pt]
& \texttt{MAG ft} ($o$)                       & & 0.188 & 0.338 & & 0.135 & 0.209 & & 0.217 & 0.311 & & 0.145 & 0.270  \\ [+5pt]

\multicolumn{7}{l}{\texttt{docT5keywords greedy decoding + keyFilT5r}} \\
\rowfont{\scriptsize}
& & & \multicolumn{1}{R}{$h$} & \multicolumn{1}{R}{$h$} & & \multicolumn{1}{R}{$ghijklmno$} & \multicolumn{1}{R}{$h$} & & \multicolumn{1}{R}{$ghijkmno$} & \multicolumn{1}{R}{$ghi$} & & \multicolumn{1}{R}{$ghijklmno$} & \multicolumn{1}{R}{$h$} \\[-5pt]
& \texttt{vanilla} ($p$)                    & & 0.186 & 0.330 & & \textBF{0.160} & 0.196 & & 0.239 & 0.266 & & \textBF{0.194} & 0.261 \\
\rowfont{\scriptsize}
& & & \multicolumn{1}{R}{$h$} & \multicolumn{1}{R}{$h$} & & \multicolumn{1}{R}{$ghijklmno$} & \multicolumn{1}{R}{$h$} & & \multicolumn{1}{R}{$ghijklmnor$} & \multicolumn{1}{R}{$ghipr$} & & \multicolumn{1}{R}{$ghijklmno$} & \multicolumn{1}{R}{$h$} \\[-5pt]
& \texttt{FlanT5} ($q$)                     & & 0.187 & 0.332 & & \textBF{0.160} & 0.195 & & \textBF{0.243} & 0.281 & & \textBF{0.194} & 0.258 \\
\rowfont{\scriptsize}
& & & \multicolumn{1}{R}{$h$} & \multicolumn{1}{R}{$h$} & & \multicolumn{1}{R}{$ghijklmno$} & \multicolumn{1}{R}{$h$} & & \multicolumn{1}{R}{$ghikmno$} & \multicolumn{1}{R}{$ghi$} & & \multicolumn{1}{R}{$ghijklmno$} & \multicolumn{1}{R}{$h$} \\[-5pt]
& \texttt{MAG ft} ($r$)                     & & 0.184 & 0.323 & & 0.159 & 0.195 & & 0.237 & 0.268 & & \textBF{0.194} & 0.260 \\

\bottomrule
\end{tabular}
\end{table}

\begin{table}
\centering
\caption{Present keyphrase labelling partial match evaluation results. Statistically significant improvements according to the Randomized Tukey HSD test (1M permutations, $\alpha = 0.05$) are shown with superscripted letters from $a$ to $r$.}
\label{table:present_partial_results}
\renewcommand{\arraystretch}{1.49}
\setlength{\tabcolsep}{5pt}
\footnotesize
\begin{tabular}
{l@{\hspace{1\tabcolsep}}rrrrrrrrrrrrr}
\toprule
\multicolumn{2}{l}{\multirow{2}*{Method}} & & \multicolumn{2}{c}{Inspec} & & \multicolumn{2}{c}{KP20k} & & \multicolumn{2}{c}{KPTimes} & & \multicolumn{2}{c}{KP-BioMed} \\
 & & & \multicolumn{1}{c} {F$_1$@5} & \multicolumn{1}{c} {F$_1$@$\mathcal{O}$} &
 & \multicolumn{1}{c} {F$_1$@5} & \multicolumn{1}{c} {F$_1$@$\mathcal{O}$} &
 & \multicolumn{1}{c} {F$_1$@5} & \multicolumn{1}{c} {F$_1$@$\mathcal{O}$} &
 & \multicolumn{1}{c} {F$_1$@5} & \multicolumn{1}{c} {F$_1$@$\mathcal{O}$}  \\

\cline{1-2} \cline{4-5} \cline{7-8} \cline{10-11} \cline{13-14}

\rowfont{\tiny}
& & & \multicolumn{1}{R}{$dfgh$} & \multicolumn{1}{R}{$dfgh$} & & \multicolumn{1}{R}{$d$} & \multicolumn{1}{R}{$d$} & & \multicolumn{1}{R}{$gh$} & \multicolumn{1}{R}{$gh$} & &  &  \\[-5pt]
\multicolumn{2}{l}{\texttt{TextRank} ($a$)}   & & 0.436 & 0.517 & & 0.315 & 0.385 & & 0.257 & 0.329 & & 0.268 & 0.292 \\
\rowfont{\tiny}
& & & \multicolumn{1}{R}{$dfgh$} & \multicolumn{1}{R}{$dfgh$} & & \multicolumn{1}{R}{$ade$} & \multicolumn{1}{R}{$ade$} & & \multicolumn{1}{R}{$agh$} & \multicolumn{1}{R}{$adgh$} & & \multicolumn{1}{R}{$a$} & \multicolumn{1}{R}{$ae$} \\[-5pt]
\multicolumn{2}{l}{\texttt{SingleRank} ($b$)} & & 0.436 & 0.506 & & 0.334 & 0.415 & & 0.284 & 0.365 & & 0.297 & 0.331 \\
\rowfont{\tiny}
& & & \multicolumn{1}{R}{$dfgh$} & \multicolumn{1}{R}{$dfgh$} & & \multicolumn{1}{R}{$abde$} & \multicolumn{1}{R}{$ade$} & & \multicolumn{1}{R}{$abdegh$} & \multicolumn{1}{R}{$abdegh$} & & \multicolumn{1}{R}{$abdegh$} & \multicolumn{1}{R}{$abdegh$} \\[-5pt]
\multicolumn{2}{l}{\texttt{TopicRank} ($c$)}  & & 0.427 & 0.480 & & 0.344 & 0.406 & & 0.336 & 0.403 & & 0.344 & 0.371 \\
\rowfont{\tiny}
& & &  & \multicolumn{1}{R}{$h$} & &  &  & & \multicolumn{1}{R}{$agh$} & \multicolumn{1}{R}{$agh$} & & \multicolumn{1}{R}{$abe$} & \multicolumn{1}{R}{$abeg$} \\[-5pt]
\multicolumn{2}{l}{\texttt{YAKE} ($d$)}       & & 0.332 & 0.375 & & 0.289 & 0.344 & & 0.284 & 0.341 & & 0.319 & 0.348 \\
\rowfont{\tiny}
& & & \multicolumn{1}{R}{$dfgh$} & \multicolumn{1}{R}{$cdfgh$} & & \multicolumn{1}{R}{$ad$} & \multicolumn{1}{R}{$ad$} & & \multicolumn{1}{R}{$agh$} & \multicolumn{1}{R}{$adgh$} & & \multicolumn{1}{R}{$a$} & \multicolumn{1}{R}{$a$} \\[-5pt]
\multicolumn{2}{l}{\texttt{EmbedRank} ($e$)}  & & 0.450 & 0.542 & & 0.323 & 0.395 & & 0.289 & 0.370 & & 0.292 & 0.321 \\
\rowfont{\tiny}
& & & \multicolumn{1}{R}{$dh$} & \multicolumn{1}{R}{$h$} & & \multicolumn{1}{R}{$abde$} & \multicolumn{1}{R}{$abcdeh$} & & \multicolumn{1}{R}{$abcdegh$} & \multicolumn{1}{R}{$abcdegh$} & & \multicolumn{1}{R}{$abcdegh$} & \multicolumn{1}{R}{$abcdegh$} \\[-5pt]
\multicolumn{2}{l}{\texttt{KEA} ($f$)}        & & 0.377 & 0.405 & & 0.348 & 0.432 & & 0.362 & 0.439 & & 0.356 & 0.390 \\
\rowfont{\tiny}
& & & \multicolumn{1}{R}{$h$} & \multicolumn{1}{R}{$h$} & & \multicolumn{1}{R}{$abcdef$} & \multicolumn{1}{R}{$abcdeh$} & & \multicolumn{1}{R}{$h$} & \multicolumn{1}{R}{$h$} & & \multicolumn{1}{R}{$abe$} & \multicolumn{1}{R}{$abe$} \\[-5pt]
\multicolumn{2}{l}{\texttt{One2Set} ($g$)}    & & 0.368 & 0.379 & & 0.361 & 0.438 & & 0.126 & 0.173 & & 0.315 & 0.340 \\
\rowfont{\tiny}
& & &  &  & & \multicolumn{1}{R}{$abcdefg$} & \multicolumn{1}{R}{$ade$} & &  &  & & \multicolumn{1}{R}{$abdeg$} & \multicolumn{1}{R}{$abe$} \\[-5pt]
\multicolumn{2}{l}{\texttt{KeyBART} ($h$)}    & & 0.299 & 0.309 & & 0.375 & 0.405 & & 0.031 & 0.036 & & 0.334 & 0.348 \\
\rowfont{\tiny}
& & & \multicolumn{1}{R}{$abcdefgh$} & \multicolumn{1}{R}{$abcdfgh$} & & \multicolumn{1}{R}{$abcdefgh$} & \multicolumn{1}{R}{$abcdefghq$} & & \multicolumn{1}{R}{$adgh$} & \multicolumn{1}{R}{$abcdefghpqr$} & & \multicolumn{1}{R}{$abcdefghq$} & \multicolumn{1}{R}{$abcdefghpqr$} \\[-5pt]
\multicolumn{2}{l}{\texttt{Vicuna} ($i$)}     & & 0.489 & 0.579 & & 0.420 & 0.497 & & 0.406 & 0.496 & & 0.463 & 0.510 \\

\midrule
\multicolumn{7}{l}{\texttt{docT5keywords greedy decoding}} \\
\rowfont{\tiny}
& & & \multicolumn{1}{R}{$abcdefghi$} & \multicolumn{1}{R}{$abcdefghi$} & & \multicolumn{1}{R}{$abcdefghikmno$} & \multicolumn{1}{R}{$abcdefghipqr$} & & \multicolumn{1}{R}{$abcdefghikmnopqr$} & \multicolumn{1}{R}{$abcdefghinpqr$} & & \multicolumn{1}{R}{$abcdefghiknpqr$} & \multicolumn{1}{R}{$abcdefghiknpqr$} \\[-5pt]
& \texttt{vanilla} ($j$)                      & & 0.543 & 0.633 & & 0.474 & 0.521 & & \textBF{0.460} & 0.521 & & \textBF{0.489} & \textBF{0.532} \\
\rowfont{\tiny}
& & & \multicolumn{1}{R}{$abcdefghi$} & \multicolumn{1}{R}{$abcdefghi$} & & \multicolumn{1}{R}{$abcdefghimno$} & \multicolumn{1}{R}{$abcdefghipqr$} & & \multicolumn{1}{R}{$abcdefghimno$} & \multicolumn{1}{R}{$abcdefghinpqr$} & & \multicolumn{1}{R}{$abcdefghiqr$} & \multicolumn{1}{R}{$abcdefghpqr$} \\[-5pt]
& \texttt{FlanT5} ($k$)                       & & 0.542 & 0.647 & & 0.465 & 0.521 & & 0.437 & \textBF{0.522} & & 0.472 & 0.518 \\
\rowfont{\tiny}
& & & \multicolumn{1}{R}{$abcdefghi$} & \multicolumn{1}{R}{$abcdefghi$} & & \multicolumn{1}{R}{$abcdefghijkmnoq$} & \multicolumn{1}{R}{$abcdefghipqr$} & & \multicolumn{1}{R}{$abcdefghikmnopqr$} & \multicolumn{1}{R}{$abcdefghinpqr$} & & \multicolumn{1}{R}{$abcdefghiknpqr$} & \multicolumn{1}{R}{$abcdefghiknpqr$} \\[-5pt]
& \texttt{MAG ft} ($l$)                       & & 0.543 & 0.634 & & \textBF{0.483} & 0.529 & & 0.458 & 0.519 & & 0.487 & 0.528 \\ [+5pt]

\multicolumn{7}{l}{\texttt{docT5keywords majority voting}} \\
\rowfont{\tiny}
& & & \multicolumn{1}{R}{$abcdefghi$} & \multicolumn{1}{R}{$abcdefghi$} & & \multicolumn{1}{R}{$abcdefghi$} & \multicolumn{1}{R}{$abcdefghipqr$} & & \multicolumn{1}{R}{$abcdefghi$} & \multicolumn{1}{R}{$abcdefghipqr$} & & \multicolumn{1}{R}{$abcdefghiknpqr$} & \multicolumn{1}{R}{$abcdefghiknpqr$} \\[-5pt]
& \texttt{vanilla} ($m$)                      & & 0.542 & 0.651 & & 0.437 & 0.529 & & 0.425 & 0.514 & & 0.487 & 0.528 \\
\rowfont{\tiny}
& & & \multicolumn{1}{R}{$abcdefghi$} & \multicolumn{1}{R}{$abcdefghi$} & & \multicolumn{1}{R}{$abcdefghi$} & \multicolumn{1}{R}{$abcdefghipqr$} & & \multicolumn{1}{R}{$abcdefghi$} & \multicolumn{1}{R}{$abcdefghipqr$} & & \multicolumn{1}{R}{$abcdefghiqr$} & \multicolumn{1}{R}{$abcdefghpqr$} \\[-5pt]
& \texttt{FlanT5} ($n$)                       & & 0.536 & 0.651 & & 0.434 & 0.528 & & 0.419 & 0.509 & & 0.473 & 0.517 \\
\rowfont{\tiny}
& & & \multicolumn{1}{R}{$abcdefghi$} & \multicolumn{1}{R}{$abcdefghi$} & & \multicolumn{1}{R}{$abcdefghin$} & \multicolumn{1}{R}{$abcdefghijkpqr$} & & \multicolumn{1}{R}{$abcdefghi$} & \multicolumn{1}{R}{$abcdefghipqr$} & & \multicolumn{1}{R}{$abcdefghiknpqr$} & \multicolumn{1}{R}{$abcdefghipqr$} \\[-5pt]
& \texttt{MAG ft} ($o$)                       & & 0.537 & 0.653 & & 0.442 & \textBF{0.535} & & 0.424 & 0.514 & & 0.486 & 0.525 \\ [+5pt]

\multicolumn{7}{l}{\texttt{docT5keywords greedy decoding + keyFilT5r}} \\
\rowfont{\tiny}
& & & \multicolumn{1}{R}{$abcdefghi$} & \multicolumn{1}{R}{$abcdefghi$} & & \multicolumn{1}{R}{$abcdefghikmno$} & \multicolumn{1}{R}{$abcdefgh$} & & \multicolumn{1}{R}{$abcdefghino$} & \multicolumn{1}{R}{$abcdegh$} & & \multicolumn{1}{R}{$abcdefghq$} & \multicolumn{1}{R}{$abcdefghq$} \\[-5pt]
& \texttt{vanilla} ($p$)                    & & 0.560 & 0.640 & & 0.479 & 0.489 & & 0.432 & 0.445 & & 0.467 & 0.471 \\
\rowfont{\tiny}
& & & \multicolumn{1}{R}{$abcdefghi$} & \multicolumn{1}{R}{$abcdefghi$} & & \multicolumn{1}{R}{$abcdefghikmno$} & \multicolumn{1}{R}{$abcdefgh$} & & \multicolumn{1}{R}{$abcdefghimno$} & \multicolumn{1}{R}{$abcdefghpr$} & & \multicolumn{1}{R}{$abcdefgh$} & \multicolumn{1}{R}{$abcdefgh$} \\[-5pt]
& \texttt{FlanT5} ($q$)                     & & \textBF{0.565} & \textBF{0.657} & & 0.474 & 0.486 & & 0.436 & 0.460 & & 0.455 & 0.458 \\
\rowfont{\tiny}
& & & \multicolumn{1}{R}{$abcdefghi$} & \multicolumn{1}{R}{$abcdefghi$} & & \multicolumn{1}{R}{$abcdefghijkmno$} & \multicolumn{1}{R}{$abcdefgh$} & & \multicolumn{1}{R}{$abcdefghimno$} & \multicolumn{1}{R}{$abcdegh$} & & \multicolumn{1}{R}{$abcdefghq$} & \multicolumn{1}{R}{$abcdefghq$} \\[-5pt]
& \texttt{MAG ft} ($r$)                     & & 0.558 & 0.632 & & 0.481 & 0.492 & & 0.434 & 0.447 & & 0.463 & 0.467 \\

\bottomrule
\end{tabular}
\end{table}

\clearpage

\section{Vicuna Guidelines}
\label{appendix:vicuna}

{ \scriptsize
\begin{lstlisting}[caption=Vicuna keyphrase generation guidelines.]
System instructions: You are an expert analyst that extracts important keywords from documents.

Input: Could you produce a list of lowercased keywords for the following documents? Please, avoid keywords with overlapping terms among them.

In today's society, people's lives are increasingly inseparable from computer information. Due to the continuous improvement of technology and the rapid development of internet technology, the network environment is becoming more and more complex, which makes it easy to cause loopholes in the information retrieval system when people use the network. Therefore, it is especially important to search for legal knowledge by computer. In order to adapt to this change and demand, we need a retrieval system to provide the corresponding search function, legal information content, and management and other services, so as to achieve the purpose of computer legal information retrieval. The legal information retrieval system is computer based, draws conclusions from the analysis of relevant data, and then applies them to judicial trial cases, criminal investigations, and other fields to provide a reference for relevant legal issues. The system is designed to combine computer technology with a criminal investigation and other fields, and then analyze the data to draw the corresponding conclusions. The retrieval algorithms used are mainly image and content retrieval algorithms, and image retrieval algorithms mainly use image segmentation technology, while content retrieval algorithms mainly use the cuckoo algorithm. At present, the information construction and economic and social development in China have become one of the issues of common concern and need to be solved by all countries in the world. The study of the legal information retrieval system is of great importance in the construction of information technology and the development of economic society.

Output: information-retrieval systems, information system, information storage and retrieval, humans, criminal investigation, legal retrieval, network environments, retrieval algorithms, search engines

Input: Could you produce a list of lowercased keywords for the following documents? Please, avoid keywords with overlapping terms among them.

Present day, Deaths due to heart strokes are increasing greatly day by day. Unfortunately, detecting such conditions in humans is a complex task. Handling such complex tasks can be done by using data sets. Heart strokes occurrence prediction can be done only by automation because we need to keep monitoring the heart rate. MITBIH arrhythmia is one of the datasets which help us, so it is used in this paper. Automation for the mentioned task can be obtained by using various data mining techniques. Some of the techniques used in this paper are decision trees, naive bayes, ANN algorithm & Random Forest algorithm. Therefore, the motto of this paper is to compare the above-mentioned algorithms & find out which is more accurate in accomplishing the task. At the end, after all the assessments we can say that the algorithm Random Forest has got 99% of accuracy which is recorded as highest among all the algorihms. But in the case of ECG images ANN algorithm has achieved an accuracy of 94%.

Output: naive bayes, cerebrovascular accident, electrocardiogram, image analysis, predictive value, ann, data mining, article, random forest, machine learning
\end{lstlisting}
}

\printcredits

\section*{Funding Sources}
This work has received support from projects: PLEC2021-007662 (MCIN/AEI/10.13039/501100011033 Ministerio de Ciencia e Innovación, European Union NextGenerationEU/PRTR) and PID2022-137061OB-C21 (MCIN/AEI/10 .13039/501100011033/, Ministerio de Ciencia e Innovación, ERDF A way of making Europe, by the European Union); Consellería de Educación, Universidade e Formación Profesional, Spain (accreditations 2019–2022 ED431G/01 and GPC ED431B 2022/33) and the European Regional Development Fund, which acknowledges the CITIC Research Center. The first author also acknowledges the support of grant DIN2020-011582 financed by the MCIN/AEI/10.13039/501100011033.

\clearpage

\bibliographystyle{model1-num-names}

\end{document}